\documentclass[onecolumn,notitlepage,nofootinbib,superscriptaddress]{revtex4-1}

\usepackage{amsmath,amssymb,bbm,soul}
\usepackage{amsthm}
\usepackage{caption}
\usepackage{subcaption}
\usepackage{graphbox,graphicx}

\catcode`\|=\active \def|{
\fontencoding{T1}\selectfont\symbol{124}\fontencoding{\encodingdefault}}

\usepackage{xcolor}

\newcommand{\id}{\mathbbm{1}} 

\usepackage[normalem]{ulem}
\newcommand{\stkout}[1]{\ifmmode\text{\sout{\ensuremath{#1}}}\else\sout{#1}\fi}

\newcommand{\ket}[1]{|#1\rangle} 

\usepackage{MnSymbol}

\begin{document}

\author{Nina Megier}

\email{nina.megier@mi.infn.it}

\affiliation{Dipartimento di Fisica “Aldo Pontremoli”, Università degli Studi di Milano, via Celoria 16, 20133 Milan, Italy}
\affiliation{ Istituto Nazionale di Fisica Nucleare, Sezione di Milano, via Celoria 16, 20133 Milan, Italy}

\author{Manuel Ponzi}

\affiliation{Dipartimento di Fisica “Aldo Pontremoli”, Università degli Studi di Milano, via Celoria 16, 20133 Milan, Italy}

\author{Andrea Smirne}
\author{Bassano Vacchini}

\affiliation{Dipartimento di Fisica “Aldo Pontremoli”, Università degli Studi di Milano, via Celoria 16, 20133 Milan, Italy}
\affiliation{ Istituto Nazionale di Fisica Nucleare, Sezione di Milano, via Celoria 16, 20133 Milan, Italy}

\begin{abstract}
Simple, controllable models play an important role to learn how to manipulate and control quantum resources. We focus here on quantum non-Markovianity and model the evolution of open quantum systems by quantum renewal processes. 
This class of quantum dynamics provides us with a phenomenological approach to characterise dynamics with a variety of non-Markovian
behaviours, here described in terms of
the trace distance between two reduced states. By adopting a trajectory picture for the open quantum system evolution, we analyse how non-Markovianity is influenced by the constituents defining the quantum renewal process, namely the time-continuous part of the dynamics, the type of jumps and the waiting time distributions. 
We focus not only on the mere value of the non-Markovianity measure, but also on how different features of the trace distance
evolution are altered, including times and number of revivals.
\end{abstract}

\title{Memory effects in quantum dynamics modelled by quantum renewal processes}

\maketitle

\section{Introduction}
    \label{sec:introduction}
Quantum phenomena are deemed to be the main ingredients of the next technological breakthroughs. Quantum correlations, quantum coherences and quantum non-Markovianity are the key resources 
to outperform classical protocols in many tasks, within the contexts of, e.g.,
communication \cite{PhysRevLett.69.2881,Bylicka2014}, teleportation \cite{Laine2014}, cryptography \cite{Pirandola:20}, metrology \cite{RevModPhys.89.041003} and thermodynamics \cite{Binder2018}, thus providing the pillars for future progresses in quantum technology \cite{Acin2018a}. 
Even though the developments of quantum theory already started at the beginning of the last century, a deep and thorough understanding of the above-mentioned features in view of their operational exploitation is still being developed \cite{Guhne2009a,Breuer2016a,Streltsov2017a}. This is why simple, controllable models play an important role to learn how to manipulate and control the quantum resources.

In this article, we will focus on the analysis of a Markov property in the quantum setting and on the description of a class of open quantum system dynamics featuring memory effects and allowing for a phenomenological treatment. The Markov property is a concept from the theory of classical stochastic processes, where a clear definition of Markov process can be introduced in terms of conditional probability distributions. This notion is connected with the memorylessness of the process, i.e. the fact that the future of the process is independent of its history. As stochastic processes are used to model 
reality in many different fields of research, as finance, biology, chemistry and social science, this is a highly relevant and often recurring concept \cite{TN_libero_mab22026897}. Stochastic processes naturally appear 
in the description of (open) classical systems where, at least in principle, the stochasticity
can be always traced back to the lack of knowledge on the underlying total Hamiltonian and the initial conditions \cite{TN_libero_mab22026897,VanKampen1992}. 
The extension of the classical formalism to the theory of open quantum systems is not straightforward, due to the invasive nature of the quantum measurements. As a consequence, many different, nonequivalent definitions of quantum Markov process were introduced, all of them aimed to reveal the occurrence of memory effects in quantum evolutions. In this respect, the notion of memory in the quantum realm still calls for a full physical interpretation. Some hints in this direction come from the framework of quantum thermodynamics, where, e.g., the connection between non-Markovianity and irreversible entropy production has been explored \cite{Popovic2018a,PhysRevE.99.012120}.

We will point out how the class of quantum renewal processes can be used as a phenomenological 
tool
to describe dynamics with different non-Markovian behaviours. Our study complements other approaches, whose starting point is rather a microscopic description specifying a reference total Hamiltonian. In particular, strategies aimed at controlling the non-Markovianity 
of the dynamics
have explored the manipulation of the system-environmental coupling \cite{Ma2014a,Man2015a, Man2015b}, or the modification of 
the reduced system itself \cite{Brito2015a,Franco2015a}. The possibility of delaying the 
occurrence of non-Markovianity \cite{Burgarth2021a}, and enhancing it by means of feedback control \cite{Zong2020a} has been also investigated.

 The existence of an underlying microscopical description of the evolution ensures that the reduced dynamics is indeed physical, i.e. the corresponding dynamical map $\Lambda_t$ which maps the initial reduced density operator $\rho(0)$ to a density operator at later time $t$: $\rho(t)=\Lambda_t[\rho(0)]$, is completely positive and trace preserving (CPTP)\footnote{This follows from the assumption that initially the reduced system and its environment are not correlated, i.e. the initial total state is factorised: $\rho_{tot}(0)=\rho(0)\otimes \rho_E(0)$.}. The density operator 
 yields the probability distributions in quantum physics, so that trace preservation of the dynamics keeps the correct normalisation of the probabilities, while complete positivity takes into account the possible entanglement of the system state with other, otherwise not involved, degrees of freedom \cite{breuerbook},
 ensuring that joint probabilities are properly defined. On the other hand, if one chooses a more phenomenological approach and fixes the form of the dynamical map or of the corresponding evolution equation for $\rho(t)$, the so-called master equation, 
the CPTP property of $\Lambda_t$ needs to be warranted. For Markovian semigroups this issue is well under control, i.e. one can specify the general structure of the corresponding master equation, of the so-called Gorini–Kossakowski–Sudarshan–Lindblad (GKSL) form \cite{Gorini1976,lindblad1976,PhysRevA.89.042120}, which describes any proper quantum evolution obeying a semigroup composition law. 
Its generalisation to the case of the so-called CP-divisible dynamics\footnote{The dynamical map $\Lambda_t$ is CP-divisible if the map $\Lambda_{t,s}$, defined as $\Lambda_t=\Lambda_{t,s}\Lambda_s$ is CPTP for all $0\leq s\leq t$.} obeying a more general composition law is also known \cite{Breuer2012a}.
However, a comparable result is still missing for general non-Markovian evolutions, although the topic has attracted a lot of interest \cite{PhysRevA.69.042107,PhysRevLett.101.140402,PhysRevA.70.010304,PhysRevA.71.020101,PhysRevLett.117.230401,PhysRevA.94.020103,Bassano2020,doi:10.1063/5.0036620,PhysRevA.91.042105,PhysRevA.75.022103,refId0,PhysRevE.79.041147,10.1007/978-3-030-24748-5_8,kossakowski2008}. 
Remarkably, in the case of quantum renewal processes that we analyse here CPTP of the dynamical map is guaranteed by construction. This makes this class of open quantum system dynamics a valuable 
tool for the phenomenological description of reduced dynamics.
In addition, despite their simplicity, quantum renewal processes can show a wide range of non-Markovian behaviours, which we will analyse in details in the following.

The rest of the article is organized as follows. In Sect.~\ref{sec:nm} we introduce the concept of non-Markovianity for stochastic processes. After this, we describe a possible definition of quantum non-Markovianity based on the monotonicity of the trace distance between two reduced states, which we adopt in the whole article. In Sect.~\ref{sec:qrp} we continue with the presentation of the renewal processes in the classical and the quantum domain, while 
Sect.~\ref{sec:qnmRP} is devoted to the trajectory picture of the reduced dynamics and to the description of the different 
elements that influence the non-Markovianity of the quantum renewal process: the time-continuous part of the dynamics, the type of jumps and the waiting time distributions governing the whole stochasticity of the jumps' times. In Sect.~\ref{sec:nmqrp} 
we analyse the impact on the non-Markovianity measure and, more generally, on the main features of the trace distance evolution, such as the number and instants of its revivals. Finally, we summarise our findings in the last Sect.~\ref{sec:concl}.              


\section{Memory effects in quantum dynamics}\label{sec:nm}

We say that a stochastic process $X(t)$, $t\geq 0$, taking values in a discrete set $\{x_i\}_{i\in \mathbb{N}}$ is Markov if the corresponding conditional probability distributions satisfy for any finite $n$ the following inequalities
\begin{align}
p_{1|n}(x_{n+1},t_{n+1}|x_{n},t_{n};\ldots;x_{1},t_{1})=p_{1|1}(x_{n+1},t_{n+1}|x_{n},t_{n}),
\end{align}
where the times are ordered: $t_{n+1} \geq t_{n} \geq \ldots \geq t_1\geq 0 $,
i.e., 
once we know the value $x_n$ of the stochastic process at time $t_n$, the past history prior to $t_n$
does not affect the predictions about the value of the process at any future time $t_{n+1}$.
Due to the invasive nature of quantum measurements, the extension of this definition to the quantum regime is not straightforward \cite{Vacchini_2011} and many different, non-equivalent definitions of quantum Markovianity have been introduced \cite{Rivas_2014,RevModPhys.88.021002,LI20181}. In most of them Markovianity is a property of the dynamical map $\Lambda_t$ itself, such as 
(C)P-divisibility \cite{Wolf2008a,PhysRevLett.105.050403,PhysRevA.89.042120}, the change of the volume of accessible states \cite{PhysRevA.88.020102} and monotonicity of the trace distance as a quantifier of state distinguishability \cite{PhysRevLett.103.210401, PhysRevA.81.062115};
the latter is the one we will adopt here. On the other hand, other approaches, such as the process matrix formalism \cite{PhysRevA.97.012127,Giarmatzi2021a},
ground the notion of quantum Markovianity on conditional probabilities 
associated with sequences of measurements, 
going beyond the single-time description of the open system dynamics and calling for multi-time correlations.

The definition of non-Markovianity we use here is based on the change of
distinguishability between system states, quantified in the original paper \cite{BLP} by means of the 
trace distance between two reduced states in the course of the evolution. The trace distance between two quantum states $\rho$, $\sigma$ is defined as
\begin{align}
\mathcal{D}(\rho,\sigma)=\frac{1}{2}\text{Tr}|\rho-\sigma|=\frac{1}{2}\sum\limits_i |v_i|,
\end{align} 
where $v_i$ are the eigenvalues of the operator $\rho-\sigma$. 
The quantum dynamics fixed by the map $\Lambda_t$
is non-Markovian if and only if the trace distance is not a monotonous function of time, i.e.,
there exist a couple of initial states $\rho(0)$ and $\sigma(0)$ and a time $t>0$ for which
\begin{align}\label{eq:NM}
\frac{d}{dt}\mathcal{D}(\rho(t),\sigma(t))>0,
\end{align} 
where $\rho(t)=\Lambda_t\rho(0)$ and $\sigma(t)=\Lambda_t\sigma(0)$.
 \begin{center}
 \minipage{1.0\textwidth}%
    \includegraphics[width=\columnwidth]{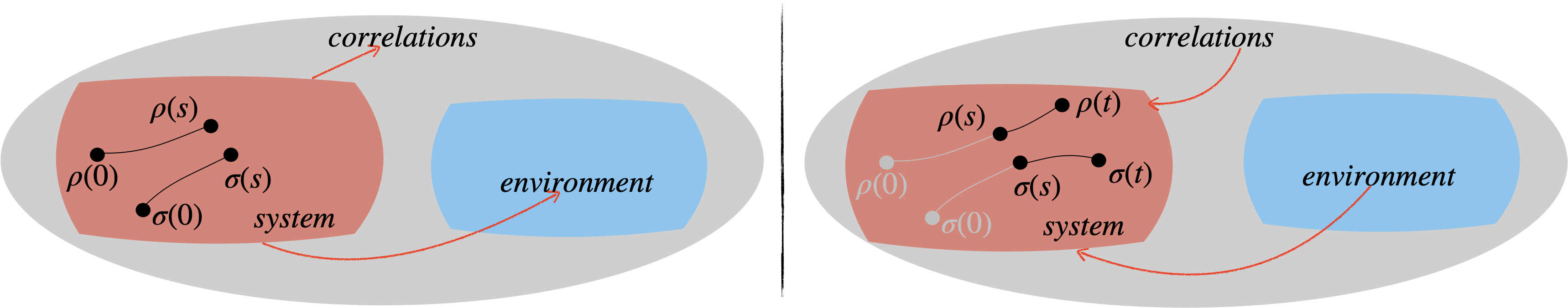}
   \endminipage
  \captionof{figure}{Sketch of the information backflow in an open quantum system dynamics, which is at the basis
   of the notion of quantum non-Markovianity used in this paper: initially the reduced states
    $\rho$, $\sigma$
    approach each other since the information is flowing out of
    the reduced system, to the environment or to the correlations between the system and the environment (left); on the other hand, an
    information backflow makes the two states
    diverge from each other at a later time (right), as can be witnessed
    via proper state distinguishability quantifiers}.
   \label{fig:InfFlow}
     \end{center}
Importantly, since the trace distance is contractive under the action of any (C)PTP map $\phi$:
\begin{align}
\mathcal{D}(\phi(\rho),\phi(\sigma))<\mathcal{D}(\rho,\sigma),
\end{align}
(C)P-divisibility \cite{Wolf2008a,PhysRevLett.105.050403,PhysRevA.89.042120,HelstromPdiv} implies monotonicity of the trace distance and thus Markovianity according to the definition above, while the converse does not hold \cite{PhysRevA.81.062115}.
The trace-distance based definition of non-Markovianity provides a clear-cut interpretation in terms of the information flow between the open quantum system and the environment as the key element possibly leading to the occurrence of memory effects in the dynamics. 
In addition, this picture allows us to trace back the exchange of information between the open system
and the environment to the correlations established by their mutual interaction \cite{Laine2010b,Mazzola2012,Smirne2013,Cialdi2014,Campbell_2019,Smirne2021}, see Fig.~\ref{fig:InfFlow}. Initially the whole information is contained in the reduced system, however, due to the system-environment interaction, some information gets transferred to external degrees of freedom in the course of the evolution. Such information can be stored both in the environment and in the system-environmental correlations. In Markovian dynamics the information flow is unidirectional, i.e. the information is always flowing from the open system to the outside world and any couple of reduced states get closer and closer with the passing of time. On the other hand, for non-Markovian evolutions some information backflow occurs, which is witnessed by an 
increase of the distance between pairs of reduced states on certain intervals of time. 
Let us stress that this viewpoint was recently strengthened, as it was shown that also different distinguishability measures between two quantum states, including entropic quantities, can be used to quantify the information flow; it appears in particular that the quantum Jensen-Shannon divergence is a natural entropic quantifier of information backflow \cite{megier2021entropic}. Additionally, a connection between monotonic contractivity of a generalisation of the trace distance and P-divisibility exists \cite{Chruscinski2011a,HelstromPdiv}, providing a common background to these approaches to non-Markovianity,
which however goes beyond the scope of this work. 

Relying on the trace distance, it is then possible to define a measure of the degree of non-Markovianity 
of a quantum dynamics. The idea is to integrate over all the revivals of the trace distance over the duration of the
dynamics, i.e., to quantify the overall amount of information flown back to the reduced system.
In addition, since we want the non-Markovianity measure to be a property of the dynamical map, while
the change of the trace distance, Eq.~\eqref{eq:NM}, will generally depend on the chosen initial
states $\rho(0)$ and $\sigma(0)$,
the non-Markovianity measure involves an optimisation over all the possible couples of initial states \cite{PhysRevLett.103.210401}:
\begin{align}\label{eq:NMm}
\mathcal{N}= \max\limits_{\rho(0), \sigma(0)} \int\limits_{d\mathcal{D}(\rho(s),\sigma(s))/ds>0} \frac{d}{ds}\mathcal{D}(\rho(s),\sigma(s))ds.
\end{align}
It was shown in \cite{PhysRevA.86.062108} that the optimal pair of states, i.e. the one achieving the maximum in the non-Markovianity measure, lies on the boundary of the states space and is made of orthogonal states. In particular, for qubit states this means that the optimal pair consists of pure states that can be represented as a pair of antipodal points on the Bloch sphere. 

\section{Renewal processes: classical and quantum}\label{sec:qrp}

Here we investigate a class of open quantum system dynamics, quantum renewal processes, which are a generalisation of a classical concept. Firstly, we briefly review semi-Markov processes, of which renewal processes are a subset, and then provide a formulation of the relevant notions in the quantum realm
\cite{Feller1968, Ross2003}.

As discussed in Sect.~\ref{sec:introduction}, the characterization of
a Markovian time evolution is essentially fixed by the GKSL theorem,
determining the expression of the generator of the dynamics. An
equivalent result for an arbitrary dynamics featuring non-Markovian
effects is not known, and only very specific results have been obtained. The
main difficulty lies in providing evolution equations whose solutions
are indeed CPTP maps. These so-called master equations can be recast in two
forms, either time local, i.e. with the functional expression \cite{breuerbook}
\begin{equation}
  \label{eq:1tcl}
  \frac{d}{dt}\rho(t)=\mathcal{L}(t)[\rho(t)],
\end{equation}
or time non-local, that is in the form \cite{nakajima,zwanzig}
\begin{equation}
  \label{eq:1nz}
  \frac{d}{dt}\rho(t)=\int\limits_{0}^t \mathcal{K}(t-s)[\rho(s)].
\end{equation}
The superoperators $\mathcal{L}(t)$ and
$\mathcal{K}(t)$ are generally related
\cite{Chruscinski2010a,Megier2020b,e22070796,Nestmann2021a},  though in a highly
non-trivial way. Moreover general conditions on their expression
warranting CPTP are not known, except for special cases. In this contribution we make reference to a large class of
well-defined evolutions obtained building on an analogy with
classical non-Markovian stochastic processes.

A semi-Markov process is a  continuous time random jump process, for which the jump probabilities are possibly site dependent but independent from each other. The probability distribution of the time between the jumps is called waiting time distribution (WTD) and provides a probability density over the positive real line
\begin{equation}
  \label{eq:1}
  f(s)\geq 0 \qquad \int\limits_0^\infty ds f(s)=1.
\end{equation} If the WTD is exponentially distributed, then the semi-Markov process reduces to a continuous time Markov chain. Otherwise, for general distributions, the memory about the time already spent in the state affects the subsequent statistics of the process, which is then non-Markovian.
The transition probabilities $T_{nm}(t)$ from the state $m$ to the state $n$ obey the equation \cite{GILLESPIE197722}
\begin{align}
   \label{eq:trans}
 \frac{d}{dt} T_{mn} (t) = \int_0^t d\tau \sum_k
 \Big[ W_{mk} (\tau) T_{kn} (t-\tau)
 - W_{km} (\tau)T_{mn} (t-\tau) \Big],
\end{align}
where the matrices $W_{mk}(t)$ are fixed by the (possibly state-dependent) WTDs $f_i(t)$,
along with the corresponding survival probabilities $g_i(t)$ defined by
\begin{align}
g_i(t)=1-\int\limits_0^t ds f_i(s),
\end{align}
and the semi-Markov matrix $\underline{\Pi}$ whose entries are the jump probabilities between sites;
in particular, denoting as ${\tilde x}(u)$ the Laplace transform of $x(t)$, one has
$\tilde{W}_{mk}(u) = { \Pi }_{mk} {\tilde f}_k(u){\tilde g}^{-1}_k(u)$.
Moreover, note that a semi-Markov process can also be seen as the merging of a renewal process and a Markovian jump process. In a renewal process, the events, here the transitions among states, occur randomly in time and the time intervals between successive events are independent. Accordingly, the evolution depends only on the current site and the time elapsed since arriving at it.
In the case of the standard renewal process all waiting times are identical, while for a so-called modified process the first $k$ waiting time distributions are different.

The notion of \textit{trajectory} is one of the basic concepts in the description of classical stochastic processes.
Indeed, in abstract terms a stochastic process can always be characterized by a suitable measure over a sample space of trajectories. Recovering a notion of trajectory is less straightforward in the quantum case, where the object of interest is the reduced density matrix $\rho(t)$,
but this can actually be done in the context of open quantum systems \cite{Barchielli_1991,breuerbook}. More specifically, the dynamics we are considering allow for an interpretation in terms of an average over trajectories in the space of operators. All the trajectories start in the same initial state $\rho(0)$, and then in each trajectory the times at which the system state undergoes discontinuous changes, the so-called \textit{jumps}, are random variables. Accordingly, the reduced density operator of the open quantum system can be obtained by a weighted sum of all possible trajectories, corresponding to fixed jump times. We will see that this point of view also helps us understand the dependence of the non-Markovianity on the specific parameters.

Quantum renewal processes are a subclass of quantum semi-Markov processes \cite{breuer2008,Vacchini_2011,Chruscinski2016a,PhysRevLett.117.230401,e22070796}, for which the time evolution reads 
\begin{eqnarray}\label{eq:semiMarkGen}
\rho(t)= p_0(t) \mathcal{F}_0(t) \rho(0)+  \sum\limits_{n=0}^\infty \int\limits_0^t dt_n\ldots \int\limits_0^{t_2} dt_1  p_n(t; t_n,\ldots,t_1) \mathcal{F}_n(t-t_n)\mathcal{E}_n \ldots.\mathcal{F}_2(t_2-t_1) \mathcal{E}_1 \mathcal{F}_1(t_1) \rho(0),
\end{eqnarray}
where the CPTP maps $\mathcal{E}_n$ describe the jumps, while the CPTP maps $\mathcal{F}_n(t)$ give the time-continuous evolutions between the jumps, and 
$p_n(t; t_n;\ldots,t_1)dt_n \ldots dt_1$ is the probability that the jumps occur (solely) around fixed times $t_1,\ldots,t_n$. Note the close analogy to the classical description recalled above. In the case of the standard process the jump times are independent and identically distributed, i.e. each waiting time has the same probability distribution and they are all mutually independent. In a modified process, instead, the probability distributions for the first jumps can differ from each other and the following ones. 

To obtain the quantum renewal processes from the general quantum semi-Markov processes one fixes the time evolution between the jumps to be of GKSL form \cite{Vacchini2020a}. What is more, one also introduces only two kinds of jumps: anterior $\mathcal{J}$ and subsequent $\mathcal{E}$ with respect to the time continuous evolution. Consequently, in quantum renewal processes one focuses on the stochastic distribution of the jumps, as in the case of classical 
renewal processes. Accordingly, we obtain the following form of the open quantum system density operator at time $t$:
\begin{eqnarray}\label{eq:RenGen}
\rho(t)= p_0(t) e^{\mathcal{L}t} \rho(0)+  \sum\limits_{n=0}^\infty \int\limits_0^t dt_n\ldots \int\limits_0^{t_2} dt_1  p_n(t; t_n,\ldots,t_1)   e^{\mathcal{M}(t-t_n)} \mathcal{E} e^{\mathcal{L}(t_n-t_{n-1})} \mathcal{J}\ldots\mathcal{E} e^{\mathcal{L}t_1} \mathcal{J}\rho(0).
\end{eqnarray}
Here, we use a ``left-ordering'', as explained in \cite{PhysRevLett.117.230401}, since a particular ordering of the operators has to be chosen in order to construct the quantum evolution from the classical counterpart. We also set in the following $\mathcal{M}=\mathcal{L}$ (the time continuous evolution is always the same) and $\mathcal{J}=\id$. With this, the above mentioned trajectories correspond to the dynamical maps $e^{\mathcal{L}(t-t_n)} \mathcal{E} e^{\mathcal{L}(t_n-t_{n-1})}\ldots\mathcal{E} e^{\mathcal{L}t_1}$, which contribute to the overall evolution with weights $p_n(t; t_n,\ldots,t_1)dt_n \ldots dt_1$.

For the standard quantum renewal process the same WTD $f(t)$ governs the whole stochasticity of the jumps' times,
\begin{eqnarray*}
p_n(t; t_n,\ldots,t_1)=g(t-t_n)\ldots f(t_2-t_1)f(t_1),
\end{eqnarray*}
where $g(t)$ is the corresponding survival probability.
When the renewal process is modified the first $k$ WTDs can be different,
\begin{align}
p_n(t; t_n,\ldots,t_1)&=g_{n+1}(t-t_n)f_{n}(t_n-t_{n-1})\ldots f_1(t_1), && n \leq k\\
p_n(t; t_n,\ldots,t_1)&=g(t-t_n)\ldots f(t_{k+1}-t_{k}) f_k(t_k-t_{k-1})\ldots f_1(t_1), && n > k.
\end{align}

Here we will investigate how the non-Markovianity of the dynamics, in terms of the monotonicity of the trace distance, depends on the choice of the involved operators, describing intermediate continuous evolutions and jumps, as well as the chosen probability distribution for the jumps. We will observe a rich variety of possible behaviours and analyse the influence of particular parameters to control the strength, time of occurrence and precise manifestation of quantum non-Markovianity.

\section{Trajectory picture}\label{sec:qnmRP}


In general there exist infinitely many different decompositions of a reduced dynamics, i.e. different ways to 
write the reduced density operator in the form
\begin{align}
\rho(t)=\sumint\limits_{i \in I} p_i(t) \rho_i(t),
\end{align}
where $I$ can be countable or uncountable set. In this representation the prefactors $p_i(t)$ can be interpreted as probabilities or probabilities densities, i.e. they are positive and normalized, and the operators $\rho_i(t)$ are themselves proper density operators, i.e. trace one and positive semi-definite. 
If the operators can be obtained by the action of CPTP maps $\Lambda^i_t$ applied on the very same initial state $\rho(0)$, each $\rho_i(t)$ can be associated to a different trajectory, whose occurrence probability is indeed given by the corresponding $p_i(t)$. 
There exist two main types of decompositions directly linked to a trajectory picture of the dynamics: time-continuous, as exemplary quantum state diffusion \cite{TN_libero_mab21458774, PhysRevA.58.1699,PhysRevLett.120.150402}, and so called jump unravelings \cite{PhysRevLett.68.580,dariusz2020open}. 
As recalled above, also quantum renewal processes have a direct decomposition in terms of trajectories, which are defined at the level of the density operators,
see in particular Eq.~\eqref{eq:RenGen}. Finally, note that an important question connected with the trajectory description of the reduced dynamics
is the existence of a continuous measurement interpretation associated with it \cite{Barchielli_1991,RevModPhys.70.101,PhysRevLett.100.080401,PhysRevLett.101.140401,PhysRevLett.124.190402,PhysRevResearch.2.043376}.

The construction of a particular trajectory can take place in two different ways. In the first method one firstly fixes the time interval $[0,T]$ 
of interest and then draws the jumps' times according to the WTDs. After each drawing if the sum of waiting times exceeds $T$ one terminates the process. 
Then the generation of the trajectory is obtained by inserting the jumps at the given times. In the second method the generation of the trajectory and drawing of the jump times take place in parallel. The time interval $[0,T]$ is divided into small intervals of length $\Delta t$, and at each intermediate midpoint one determines randomly if the jump takes place or not, with the probability fixed by the corresponding waiting time distribution.
In this second approach, fixing the  time interval $[0,T]$ in advance is in principle not necessary as one can decide along the trajectory when to stop the evolution. Note, that for a modified renewal process only the first method is applicable for the case in which the last $k$ waiting time distributions are different, a situation which was introduced in \cite{Bassano2020} under the name of inverse time operator ordering. The same is true when the last time-continuous evolution is different from the preceding ones, $\mathcal{M}\neq \mathcal{L}$ in Eq.~\eqref{eq:RenGen}, or in processes starting with a jump rather than with a time continuous evolution, $\mathcal{J}\neq \id$ in Eq.~\eqref{eq:RenGen}. In all these situations one has to know beforehand, i.e. before one starts to generate the trajectory, how many jumps occur in the investigated time interval $[0,T]$,
to know which waiting time distribution or which time evolution has to be used to generate the trajectory at a particular point of time. In this paper, for simplicity, we restrict ourselves to cases where both methods to generate the trajectory can be implemented.  We will see that the trajectory point of view in describing the evolution let us better understand the influence of the particular parameters on the non-Markovianity of the corresponding dynamical map.

The quantum renewal processes, due to the non-trivial interplay between the operatorial and stochastic contributions, can manifest a wide range of non-Markovian behaviours. However, if one assumes that all WTDs coincide, i.e. the quantum renewal process
is unmodified, and are given by an exponential distribution 
\begin{align}
f(t)=\mu e^{-\mu t},
\end{align}
where $\mu$ is the corresponding rate,
the issue simplifies significantly.
In this case, a simple connection between the WTD $f(t)$ and the associated survival probability $g(t)$ exists: $f(t)=\mu g(t)$. As shown in \cite{Vacchini2020a}, the corresponding memory kernel in the Laplace picture reads
\begin{align}
\tilde{\mathcal{K}}(u)= \mathcal{L}+ (\mathcal{E}  - 1) \tilde{f}(u-\mathcal{L}) \tilde{g}^{-1}(u-\mathcal{L}).
\end{align}
Accordingly, in time domain we obtain for this case
\begin{align}
\mathcal{K}(t)=\delta(t) [\mathcal{L}+\mu (\mathcal{E}  - 1)],
\end{align}
no matter what the generator $\mathcal{L}$ and the jump operator $\mathcal{E}$ are. This memory kernel corresponds to a quantum dynamical semigroup, and, accordingly, the underlying evolution is Markovian. 

To go beyond this case, we analyse how the time continuous dynamics, type of jumps and waiting time distributions influence qualitatively and quantitatively the non-Markovianity of the corresponding process. We focus on qubit evolutions, so that the trace distance between two quantum states equals the half of the  Euclidian distance of these states when depicted on the Bloch ball. Recall that any qubit state can be written as
\begin{align}
\rho=\frac{1}{2}(\mathbbm{1}+\vec{r} \cdot\vec{\sigma}),
\end{align}
with the vector $\vec{\sigma}$ consisting of the Pauli matrices, $\vec{\sigma}^T=(\sigma_1,\sigma_2,\sigma_3)$,
and $\vec{r}^T=(x,y,z)$ defining the Bloch vector representation of the state $\rho$.
Accordingly, the trace distance between two qubit states evolving via a quantum renewal process can be calculated as
\begin{align}
\label{td}
\mathcal{D}(\rho^1(t),\rho^2(t))=\frac{1}{2} \lim\limits_{N\rightarrow \infty}\sqrt{\left(\frac{1}{N}\sum\limits_{n=1}^N \Delta^x_n(t)\right)^2 + \left(\frac{1}{N}\sum\limits_{n=1}^N \Delta^y_n(t)\right)^2 +\left(\frac{1}{N}\sum\limits_{n=1}^N \Delta^z_n(t)\right)^2 },
\end{align}
where the sums are running over realisations of the stochastic process governed by the associated WTDs, and $\Delta^i_n(t)$ corresponds to difference of the $i$-coordinates in $n$-th realisation, e.g.
\begin{align}
\label{coord}
\Delta^x_n(t) = x^1_n(t)-  x^2_n(t),
\end{align}
which we call an $x$-trajectory. {Consequently, the trace distance between two states is not an average trace distance between the corresponding random trajectories and calculating the trace distance has to occur after generating the whole set of trajectories. 
}
Note that to have non-monotonicity in the trace distance, a non-monotonicity of the absolute value of at least one of the coordinates is necessary. This is the case not only when one of the coordinates is non-monotonous, but also when it changes its sign. This can only happen when some realisations of the trajectories include a sign change. This is however not a sufficient condition, as we will see in the following.  

We now set the different elements of the quantum renewal processes fixing the resulting trajectories and average dynamics.

\subsection{Intermediate evolutions}

We choose the time continuous evolution to be unital 
\begin{align}
\label{L}
\mathcal{L}[\rho]=\sum \limits_{k=1}^3 \frac{1}{2}\gamma_k (\sigma_k \rho \sigma_k - \rho),
\end{align}
with $\gamma_j \geq 0$ and
\begin{align}
\label{lambda}
e^{\mathcal{L}t}[\sigma_i]=e^{-t\lambda_i} \sigma_i  &&\lambda_i=\gamma_j + \gamma_k, && \text{for } i \neq j \neq k.
\end{align}
Choosing a unital dynamical map does not affect the trace distance measure of non-Markovianity, which is anyhow insensitive to translations of the Bloch sphere \cite{Rivas_2014,PhysRevA.87.042103,megier2021entropic}. As the time-continuous evolution introduced above describes a monotonic contraction of the Bloch sphere, we do not expect that it introduces any memory effects. Indeed, the dynamical map $e^{\mathcal{L}t}$ is not only Markovian according to the distinguishability criterion introduced in \cite{BLP}, but it is a CP-divisible semigroup.
We will see that a greater "strength" of this dephasing evolution - corresponding to larger values of the $\lambda$'s introduced in Eq.~\eqref{lambda} - will result in smaller non-Markovianity of the associated quantum renewal process.

\subsection{Jumps}

As said above, the quantum non-Markovianity will not occur if for all realisations of the stochastic process the coordinates, Eq.~\eqref{coord}, are monotonic and do not change sign. Accordingly, a jump channel which only consists of a contraction (and possibly translation, which, however, cannot be detected by the trace distance condition - see comment above) will necessarily lead to a Markovian dynamics. An example of such a channel is the amplitude damping (AD) channel $\mathcal{E}_{\mathrm{AD}}$, with Kraus operators
\begin{align}
K_0= \begin{pmatrix}
1  & 0\\
0 & \sqrt{1-\gamma}
\end{pmatrix},
&&
K_1= \begin{pmatrix}
0  & \sqrt{\gamma}\\
0 & 0
\end{pmatrix},
\end{align}
which shrinks the Bloch ball and translates it along the z-axes by
factors determined by the decay rate $\gamma$.
Consequently, no non-Markovianity is detected, no matter what probability distribution drives the stochasticity of the jumps' times. In particular, also for a choice of classically non-Markovian waiting time distributions, as Erlang distributions {introduced later,} one still obtains Markovian evolution, according to the trace distance criterion. 

Consequently, the next step is to choose a jump channel that results in changing the sign of the trajectories. We have chosen the $x$-Pauli channel composed with AD:
$$\mathcal{E}_{x-\mathrm{AD}}=\mathcal{E}_{x}\circ \mathcal{E}_{\mathrm{AD}},$$ 
with
\begin{align*}
\mathcal{E}_{x}[\rho]= \sigma_x \rho \sigma_x.
\end{align*}
The Pauli channels describe a $\pi$ rotation about the corresponding axis and in particular we focus here on the composition of the AD channel with the $x$-PC. {For this jump channel we will, indeed, manage to detect non-Markovianity, depending on the choice of parameters determining the dynamics.}

{Note, that, as the superoperators $\mathcal{E}_{x}$ and $\mathcal{E}_{\mathrm{AD}}$ do not commute, the jump channels $\mathcal{E}_{x-\mathrm{AD}}$ and $\mathcal{E}_{\mathrm{AD}-x}=\mathcal{E}_{\mathrm{AD}} \circ\mathcal{E}_{x}$ are different. Generally speaking, the latter possibility leads to a slightly greater non-Markovianity measure, as the jumps occur before the disruptive AD channel. Nonetheless, the qualitative behaviour for both of the choices is similar, and for simplicity we restrict here to $\mathcal{E}_{x-\mathrm{AD}}$.}

\subsection{Waiting time distributions}

{As noticed earlier, when the underlying WTDs are exponentials and the process is unmodified, i.e. all WTDs are the same, the evolution is Markovian, independently of the choice of the jump channel. This is the case even if the trajectories are non-monotonic and the sign changes take place, so, in particular, for the channel $\mathcal{E}_{x-\mathrm{AD}}$ investigated by us. However, the situation drastically changes if we allow for a modified quantum renewal process. Even if all the WTDs are exponentials, but the first $k$-th of them have different rates, we can observe a high variety of different behaviours. In particular, the number of revivals strongly depends on the choice of the rates.}

There is, however, no need to restrict our choice of WTDs to exponentials. To go beyond this case, 
we also analyse the quantum renewal process dynamics where the stochasticity of the jumps is governed by the Erlang WTD, which reads in the Laplace domain \cite{cox}\footnote{In book \cite{cox} the Erlang distribution is called the special Erlangian distribution.}:
\begin{align}
\label{erlang}
\tilde{f}_r(u)=\left(\frac{\mu}{\mu + u }\right)^r,
\end{align}  
from which one can see that it is the convolution of $r$ exponential
distributions with the same rate parameter $\mu$. The ratio
${r}/{\mu}$ fixes the mean waiting time
while the variance reads ${r}/{\mu^2}$. Accordingly, for the Erlang WTDs the mean value and the variance can be independently varied, as contrasted with the exponential WTD, where the mean waiting time $1/\mu$ fixes the variance.

We will see that in the case of Erlang WTDs even the unmodified process can lead to non-Markovianity.


   \begin{center}
   \minipage{0.45\textwidth}
             \includegraphics[align=c,width=\textwidth]{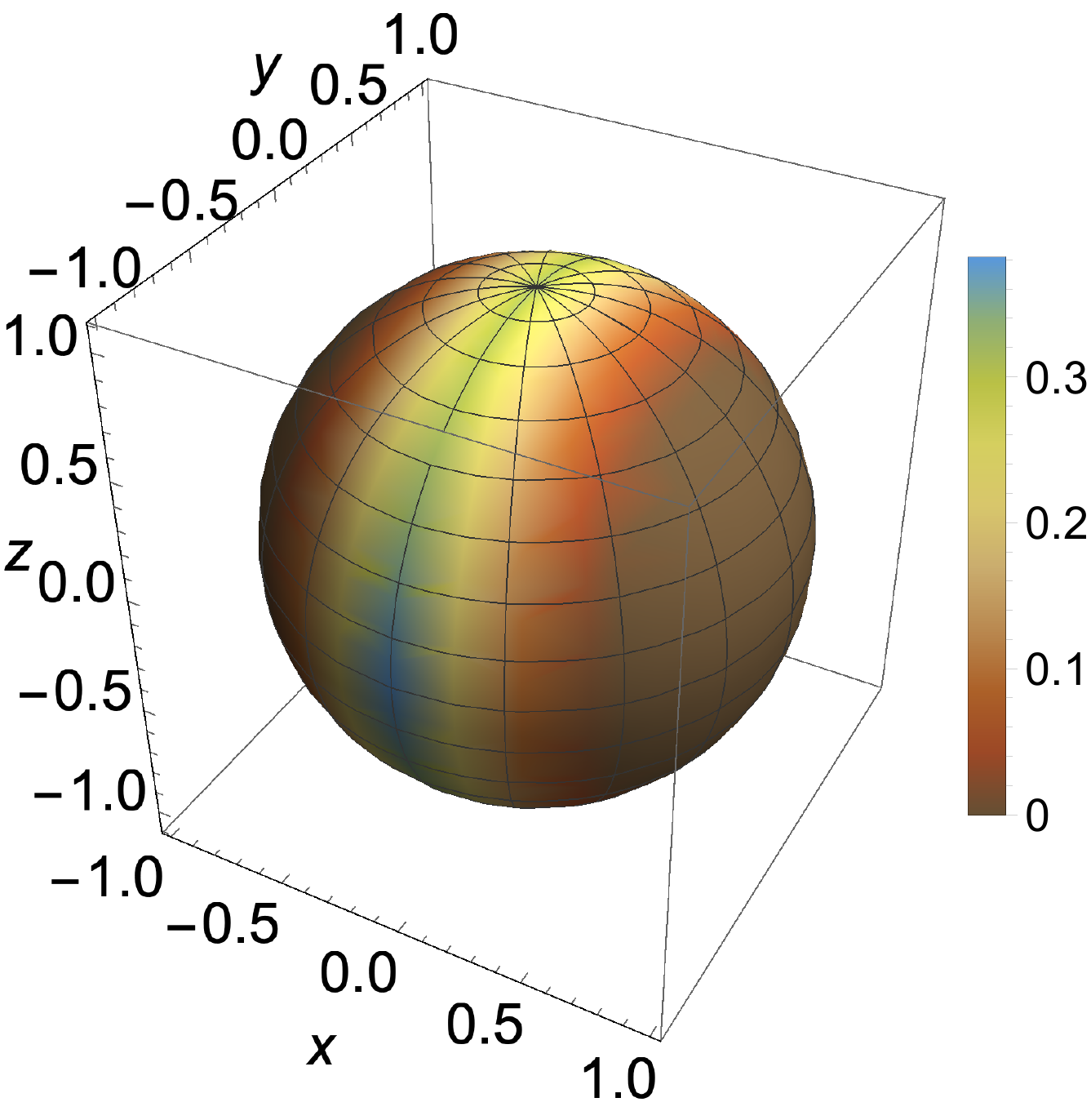}       
             \endminipage
         \hspace{1cm}
         \minipage{0.45\textwidth}
   \includegraphics[align=c,width=\textwidth]{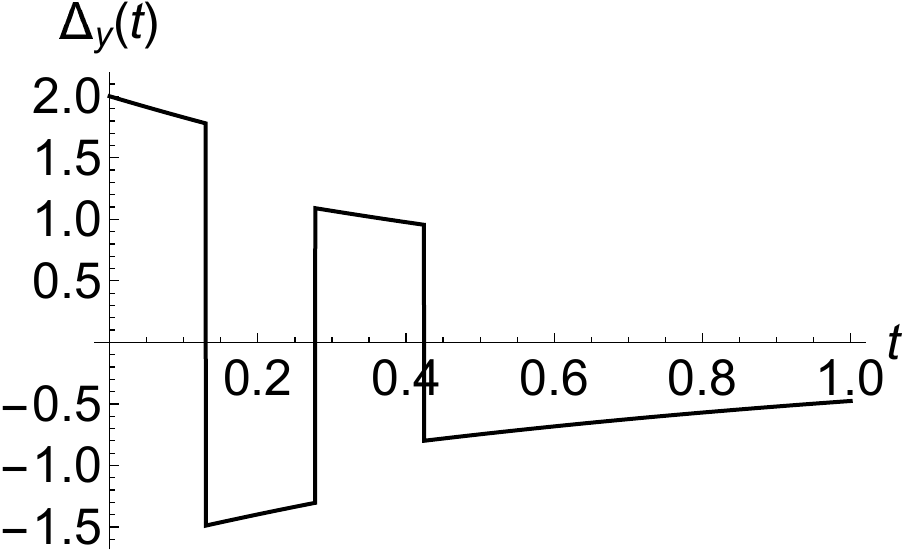}
   \endminipage
   \captionof{figure}{
   Left: Value of the non-Markovianity
    measure of a quantum renewal process as given in Eq.~\eqref{eq:NMm} in the dependence on the choice of initial orthogonal pure states, identified by the extremes of a diameter in the Bloch sphere; it clearly appears that optimal pairs lie on a vertical equator. Right: $y$-component for the trajectory in a particular realization of the process corresponding to an initial  optimal pair $\ket{\phi_{1/2}}=\frac{1}{\sqrt{2}}(\ket{0} \pm i\ket{1})$. We are here considering a $\mathcal{E}_{x-\mathrm{AD}}$ jump channel and parameters $\gamma=0.3$, $\mu=1$, $\mu_1=10$, $\lambda_{1}=\lambda_{2}=\lambda_{3}= 0.9$. Here and in the following we work in arbitrary units.
}
  \label{Fig:2}
 \end{center}

\section{Non-Markovianity of quantum renewal processes}\label{sec:nmqrp}

As mentioned in Sect.~\ref{sec:nm}, occurrence and strength of
memory effects depend on the chosen pair of initial states.
This is clarified in Fig.~\ref{Fig:2},
left, where the value of the non-Markovianity measure for the case of
the jump operator {$\mathcal{E}_{x-\mathrm{AD}}$} is plotted as a
function of the direction identifying a pair of pure orthogonal
states, corresponding to points on the Bloch sphere. It clearly appears
that the maximum is attained for states $\ket{\phi_{1/2}}=\frac{1}{\sqrt{2}}(\ket{0} \pm i\ket{1})$. We will
therefore in the following consider always this pair of initial states
lying on the $y$-axes. Note that for this choice
$\Delta_x(t)=\Delta_z(t)=0$, corresponding to the fact that
the $x$ and $z$ components of the Bloch vector of the two evolving states remain equal, so that the only relevant parameter in
the continuous time evolution is the rate
$\lambda_2$. This behaviour is due to our particular choice of the jump channel, leading to a rotation about the $x$-axis. A typical trajectory of the $y$-component
of the Bloch vector is depicted in 
Fig.~\ref{Fig:2}, right, characterized by sign changes
which determine possible non monotonicity of the trace distance obtained as in Eq.~\eqref{td}. In our analysis we will not only investigate the mere change of the non-Markovianity measure, but also the way the trace distance evolution is altered: number of revivals, times of revivals and other qualitative features.

\subsection{Exponential WTD - general results}
\label{sec:Exp1}

Here, we focus on the behaviour of the trace distance in the case of exponential WTDs. Accordingly, beside the dependence on the dephasing rate of the continuous time evolution
$\lambda_2$ and the decay rate $\gamma$ corresponding  to the strength of AD jumps, the non-Markovianity is also influenced by the rates $\mu_i$ fixing the exponential WTDs.

\begin{center}
  $\vcenter{\hbox{ 
  \includegraphics[width=.40\textwidth]{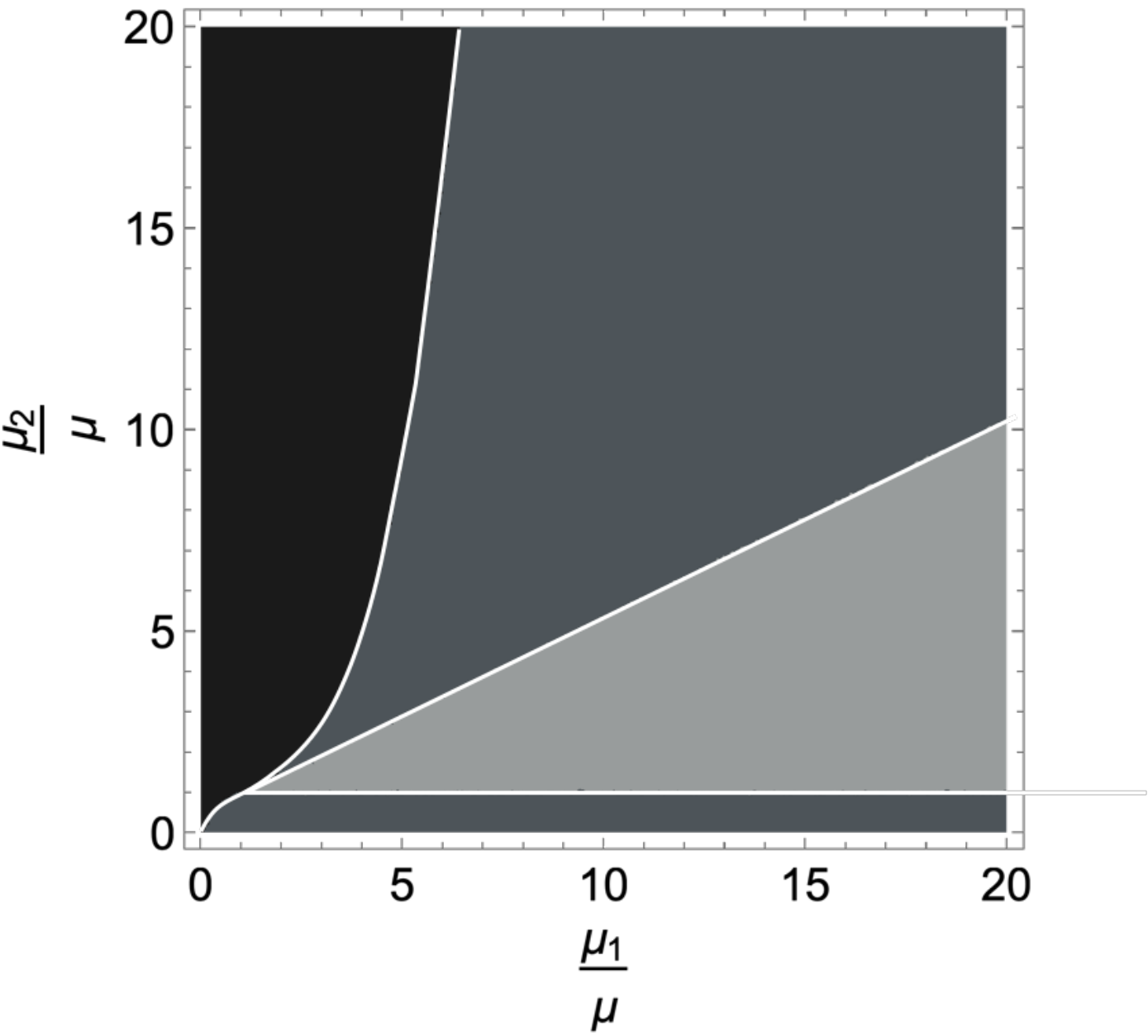}}}$
     $\vcenter{\hbox{   \includegraphics[width=.36\textwidth]{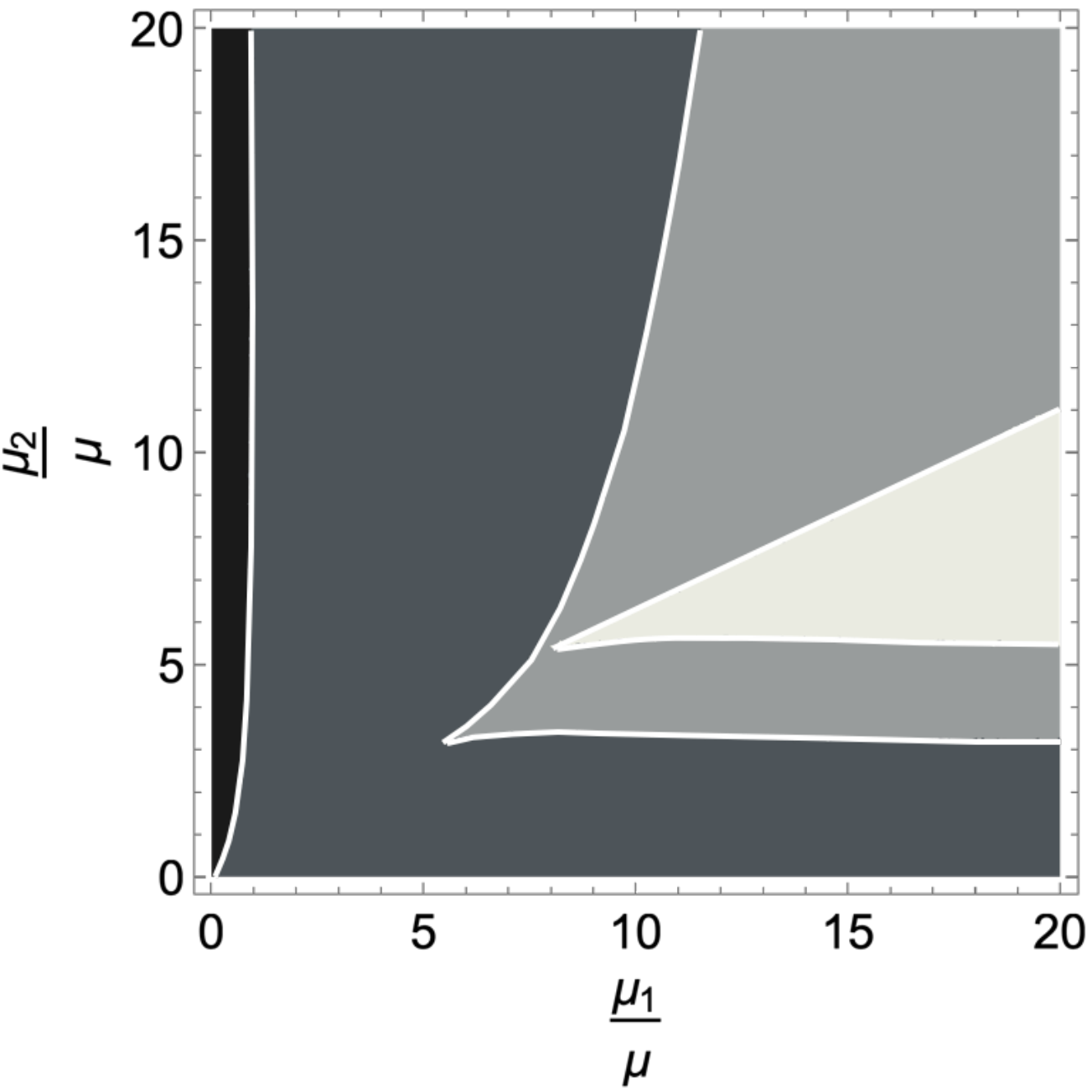}}}$
     \hspace{0.1cm}
        $\vcenter{\hbox{    \includegraphics[width=.12\textwidth]{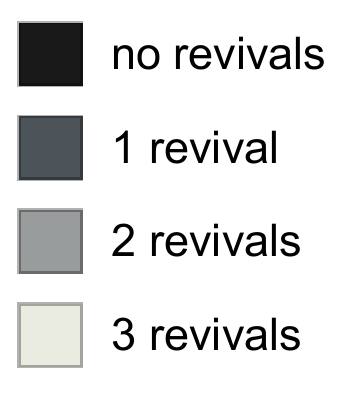}}}$
      \captionof{figure}{The number of revivals of the trace distance for a quantum renewal process with $\mathcal{E}_{x}$ jump channel
      in its dependence on the value of the rates fixing the WTDs. The maximal number of revivals for the modified quantum renewal process with $k$ exponential WTDs equals $k-1$ (here we take $k=3$ and $k=4$ from left to right; $\lambda_2=0.9$, $\mu=1$ and $\mu_3=3$ (right panel) in arbitrary units). The white lines mark the boundaries between parameter regions corresponding to processes whose trajectories exhibit different number of jumps.}
  \label{Fig:3}
\end{center}

	
	\begin{center}
   \minipage{0.31\textwidth}
      \includegraphics[width=\textwidth]{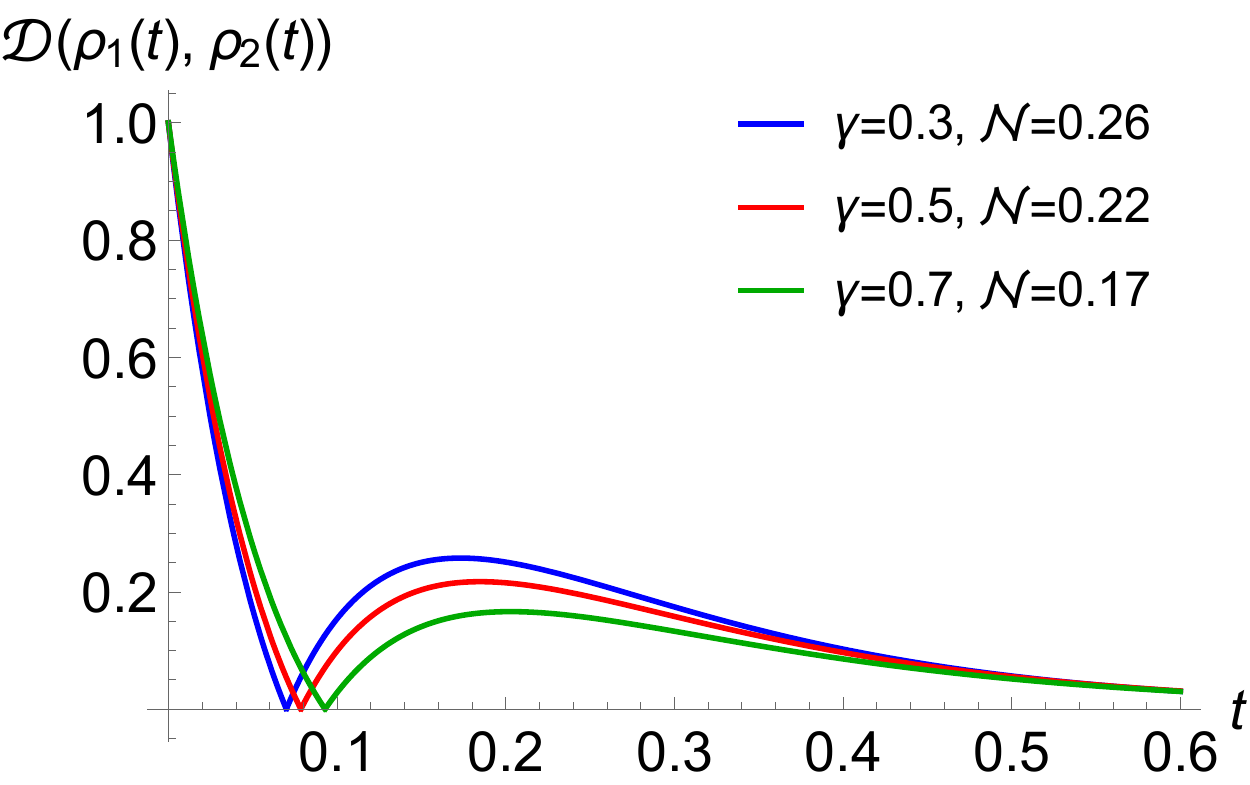}
      \endminipage
       \minipage{0.31\textwidth}
      \includegraphics[width=\textwidth]{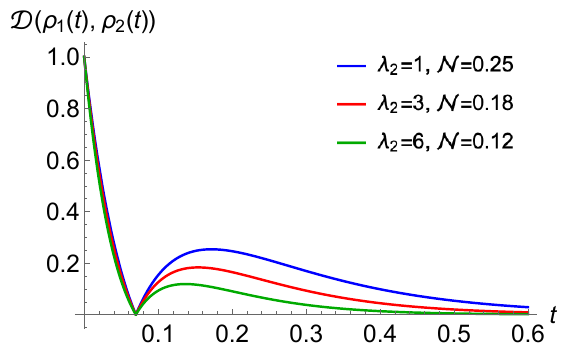} 
      \endminipage
        \minipage{0.31\textwidth}
      \includegraphics[width=\textwidth]{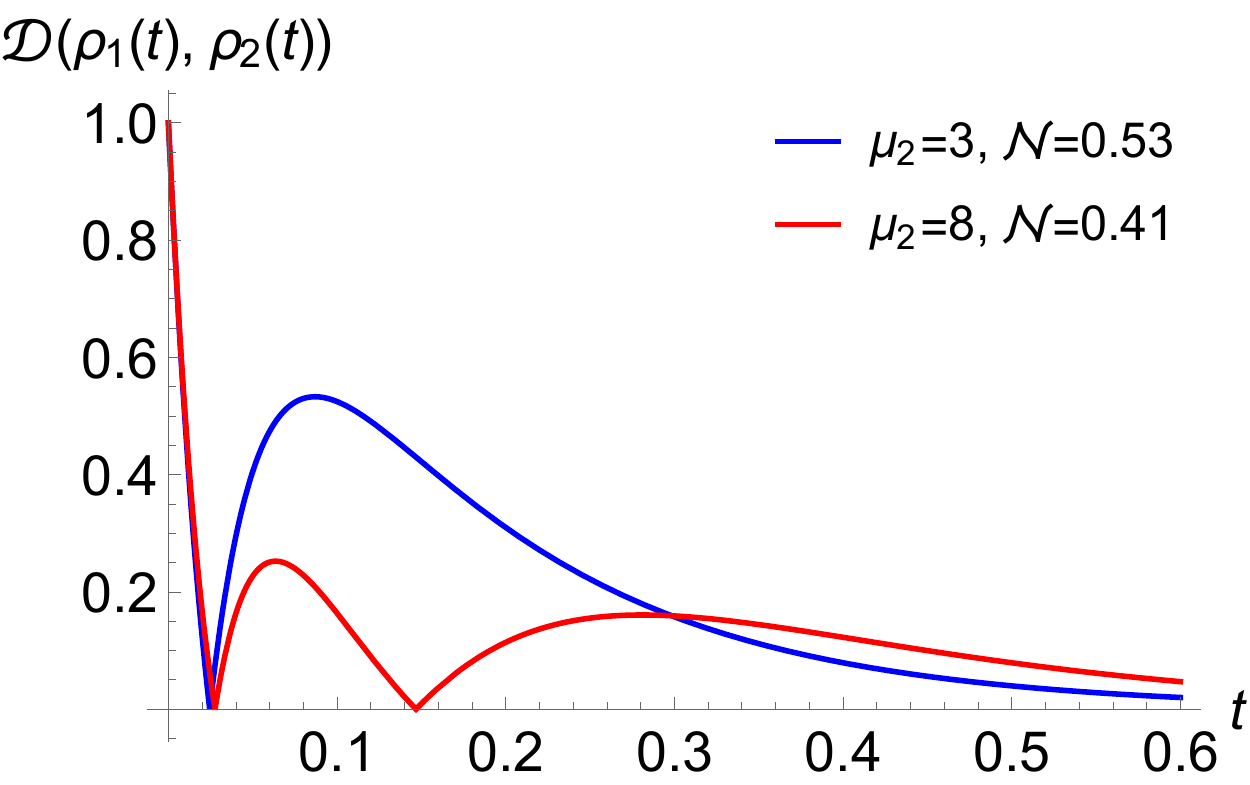}  
      \endminipage
    \captionof{figure}{
   The trace distance, testifying non-Markovianity when showing a non-monotonic behavior, for a quantum renewal process with exponential WTDs. In the left panel jumps are realized by means of a $\mathcal{E}_{x-\mathrm{AD}}$ jump channel, and one can appreaciate the reduction of the revivals for growing damping. In the middle panel jumps are given by $\mathcal{E}_{x-\mathrm{AD}}$ and stronger dephasing in the intermediate time evolution again suppresses non-Markovianity. The right panel, with jump operator $\mathcal{E}_{x}$, shows how a  larger  number  of  revivals  does  not  necessarily  lead  to  a  higher non-Markovianity measure. Across the panels $\lambda_2=0.9$, $\mu=3$ and $\mu_1=13$, apart from the last panel with $\mu_1=30$.}
  \label{Fig:4}
\end{center}

  \begin{center}
     \minipage{0.45\textwidth}
        \includegraphics[width=\textwidth]{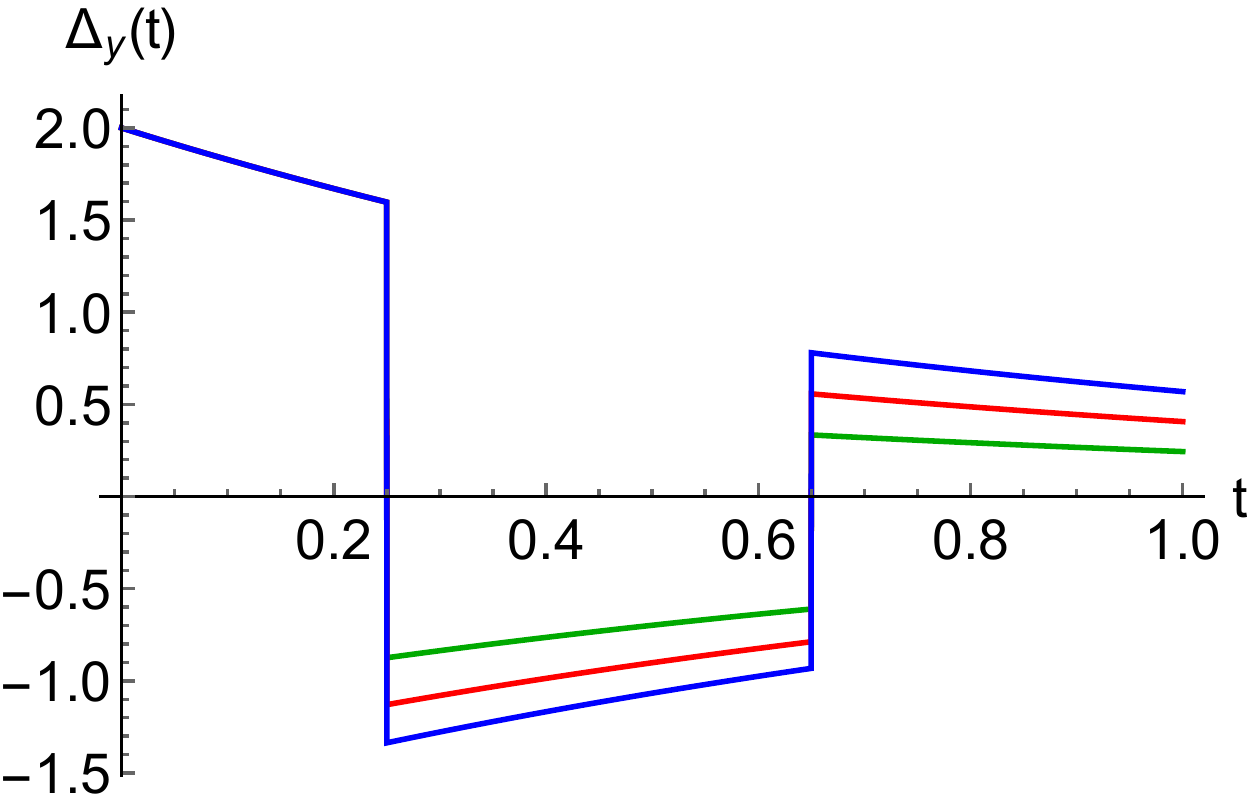}
        \endminipage
    \hspace{1cm}
       \minipage{0.45\textwidth}
    \includegraphics[width=\textwidth]{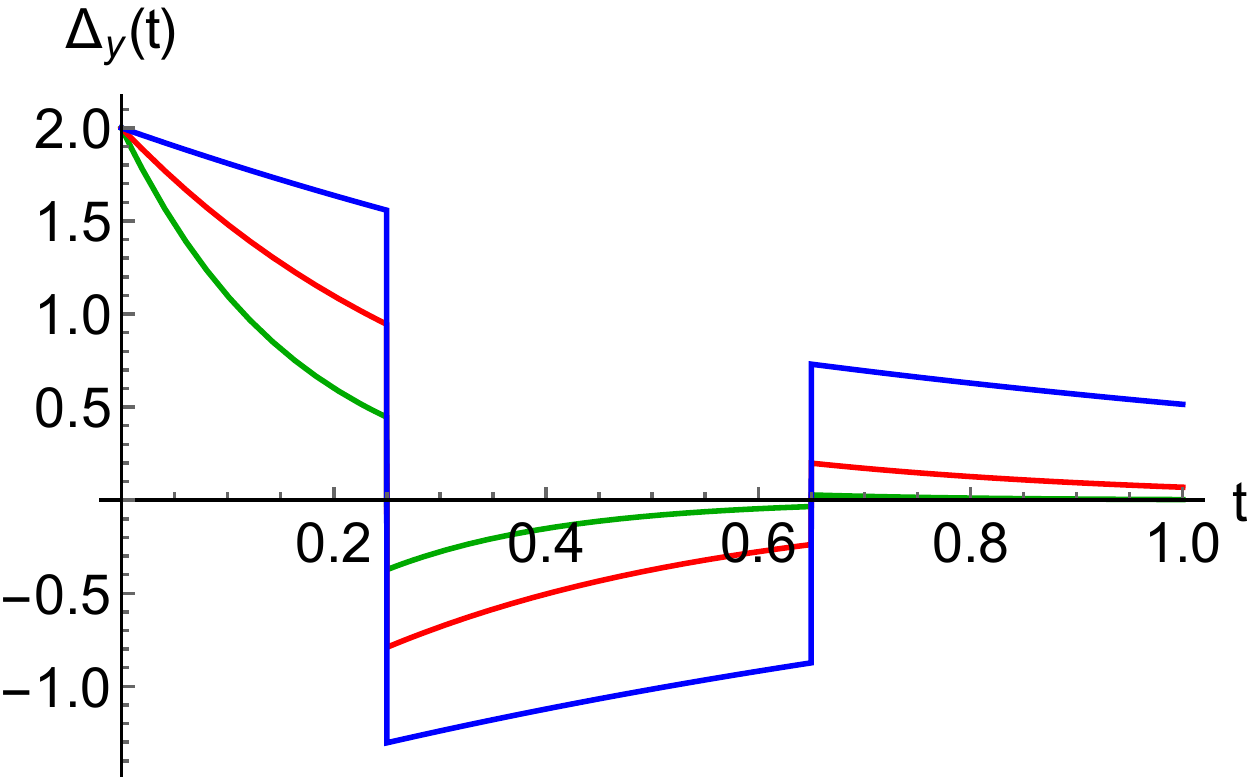}
    \endminipage
  \captionof{figure}{
  Examples of jump trajectories with  parameters as in Fig.~\ref{Fig:4}, left and middle, respectively.
  In the left panel we vary the damping rate $\gamma$, while in the right panel we vary the decay strength $\lambda_2$ associated to the intermediate time evolution.}
  \label{Fig:5}
\end{center}

The number of revivals, i.e. time intervals where the trace distance
grows, strongly depends on the number of different WTDs and on the
corresponding rates. It can be
observed that for a process with $k$-WTDs the maximal number of
revivals is $k-1$ and can only be reached if the following relation
between the rates is satisfied: 
\begin{align}
\label{relation}
\mu_1>\mu_2>\ldots>\mu_{k-1}>\mu.
\end{align}
This fact is investigated in Fig.~\ref{Fig:3}, where we report the number of revivals for a modified process with $\mathcal{E}_{x}$ jump channel and with 3 WTDs (left) or 4 WTDs (right) in dependence on the rate values. Note that throughout the manuscript we work in arbitrary units. 
The different coloured areas correspond to different numbers of
revivals, clearly growing with the number of WTDs and depending on
the corresponding rates.
The presence of amplitude damping in the jump decreases the parameter range
corresponding to higher number of revivals. At the same time the AD reduces
the value of the non-Markovianity measure. This is put into evidence
in Fig.~\ref{Fig:4}, left, where the behaviour of the trace
distance is plotted together with the estimate for the associated
non-Markovianity measure, corresponding to the sum of the revival
heights. A similar effect is obtained by
increasing the strength of the dephasing rate
$\lambda_2$ describing the time
continuous dynamics, as shown in Fig.
\ref{Fig:4}, middle, where only {$\mathcal{E}_{x}$}
determines the jumps. 

We further stress that a higher number of revivals does not
necessarily lead to a higher non-Markovianity measure, see Fig.
\ref{Fig:4}, right. Non-Markovianity is enhanced when the rate of the first WTD is much larger than the rate of the following one, {$\mu_1  \gg  \mu_2$}, allowing for a larger revival.  
Subsequent rates play a less relevant role, since, on average, the dephasing has become more effective by the time the corresponding jump occurs.

The different role of $\gamma$ and
$\lambda_2$ is visible by comparing Fig.
\ref{Fig:4}, left and Fig.
\ref{Fig:4}, middle, noticing that only
$\gamma$ affects the value of the (first) revival time.
Their different influence at the level of the trajectories is visualised in Fig.
\ref{Fig:5}. As one can observe, an increase of the decay rate implies that the height of the jumps decreases, while it does not affect the previous time continuous dynamics. This is different in the case of varying $\lambda_2$, where both the extension and the starting point of the jumps is changed and the influence on the revival time after averaging over all trajectories is wiped out. 

One can also understand the necessity of the hierarchy
given in \eqref{relation} to have the maximal
number of revivals, as well as their maximum number $k-1$. 
When the condition is satisfied, then (approximately) the first, second, \ldots, $k-1$ jumps do not influence each other. With this we mean that the $n$-th jump occurs when $n-1$ jumps have already taken place in most of the trajectories. Accordingly, the $k-1$ first jumps are connected with the revival of the trace distance, while the following jumps do not result in the revivals. The reason is that for the exponential WTDs the mean value and the variance cannot be modified independently and are such that for an unmodified process the trace distance is monotonically decreasing, as was shown in Sect.~\ref{sec:qnmRP}. This will be different in the case of Erlang WTD, which we discuss in Sects \ref{sec:Erl1} and \ref{sec:Erl2}.  When the condition \eqref{relation} is not satisfied, the number of the revivals for a modified process with $k$ different WTDs will be smaller than $k-1$.

All revivals depicted till now started when the trace distance assumed value zero, i.e. when at the
associated time the evolved states are the same. This can be seen as
a special realization of non-Markovian behaviour, since in this case the dynamical map is neither invertible nor divisible. This is, however, not always the case. We observe that a revival occurs for larger values of the trace distance when the condition $\mu_2 \gtrapprox \mu_1 \gg \mu$ (3-WTDs process) is satisfied, see Fig.~\ref{1}, left, where the $\mu_2$ is varied, and right, where $\mu$ is altered. The mean waiting time of the second jump is small
enough with respect to the first jump to prevent the trace distance to reach zero, and the following third jumps occur too late to change this tendency. Note, that in this case the maximal number of revivals, $k-1$, cannot be reached. 

  \begin{center}
       \minipage{0.47\textwidth}
    \includegraphics[width=\textwidth]{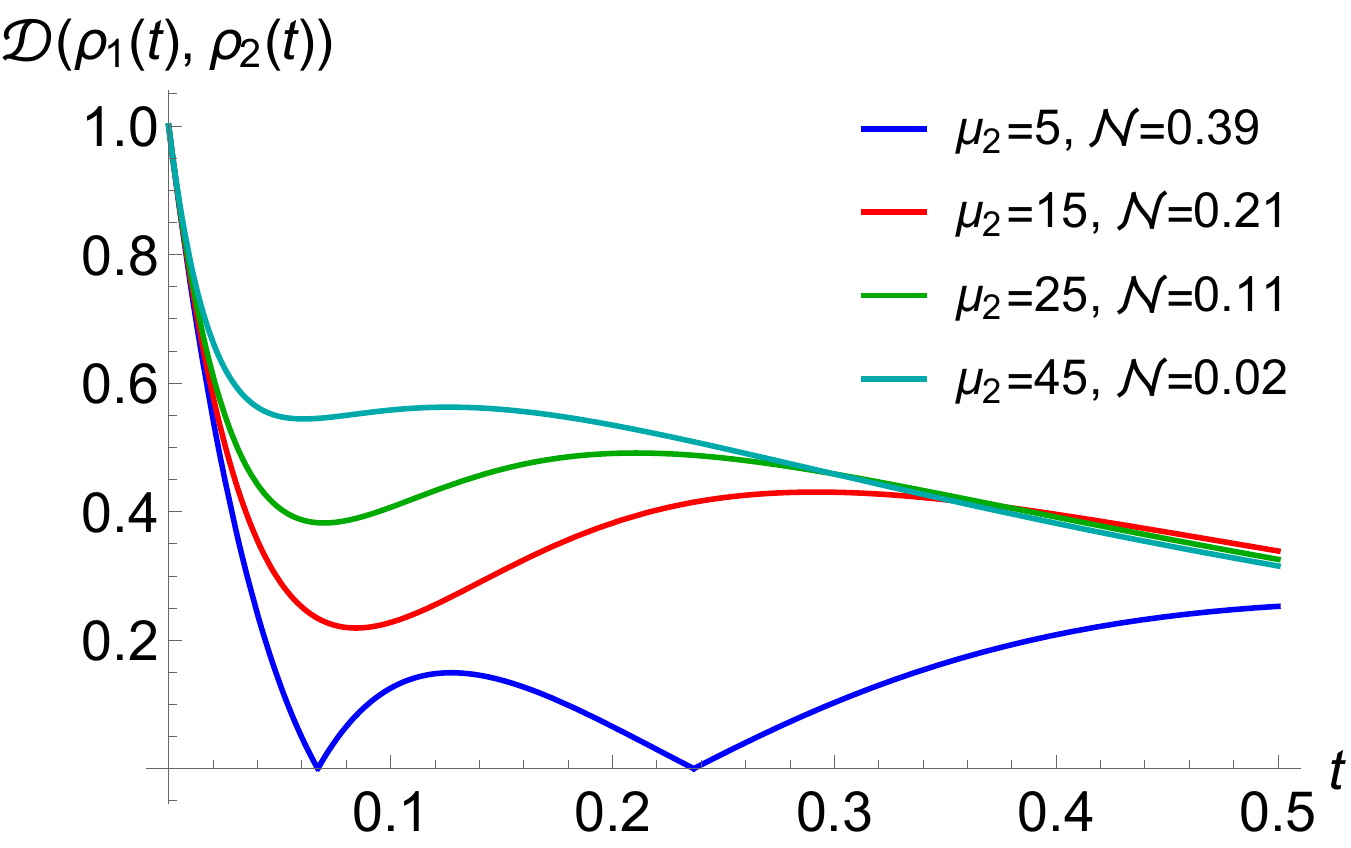}
    \endminipage
    \hspace{1cm}
         \minipage{0.45\textwidth}
      \includegraphics[width=\textwidth]{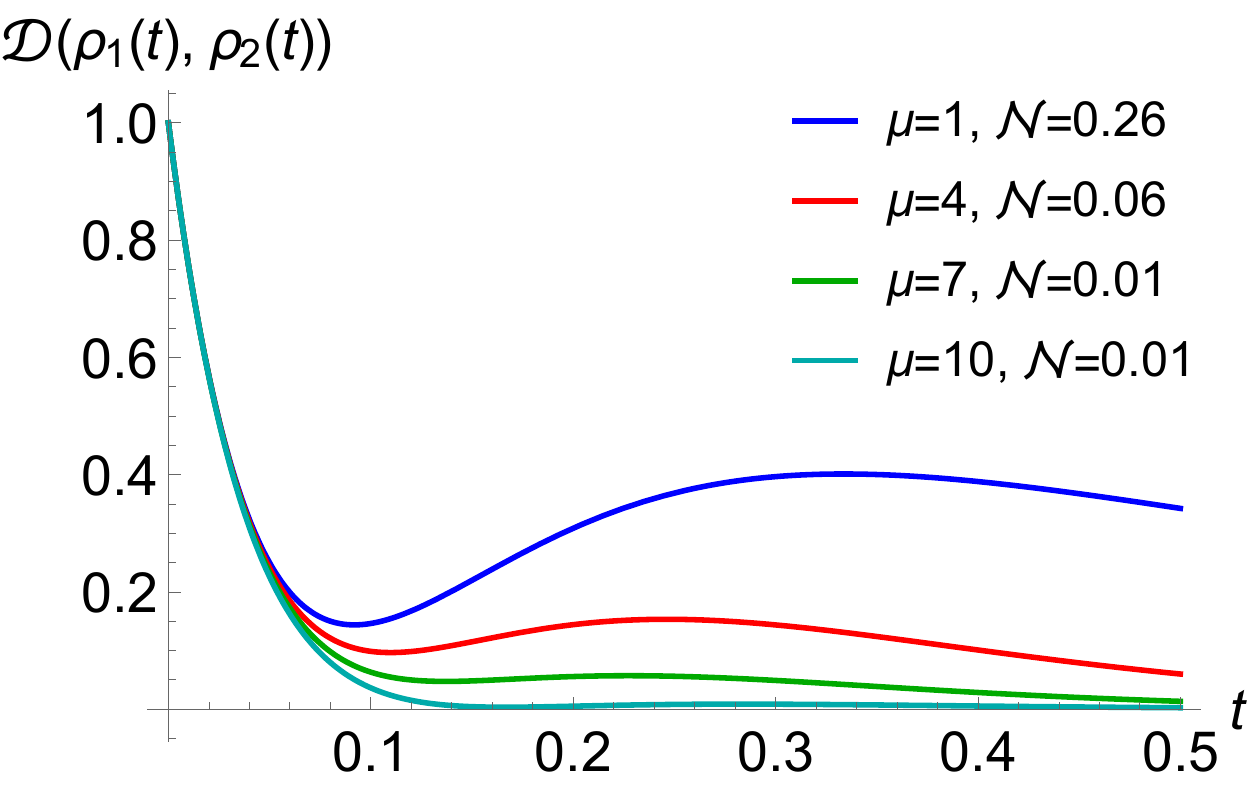} 
      \endminipage
  \captionof{figure}{The plots show the dependence of the times at which revivals take place on the rates of the exponential WTDs.
  We consider the $\mathcal{E}_{x-\mathrm{AD}}$ jump channel together with damping rate $\gamma=0.3$ and relaxation rate $\lambda_2=0.1$. Overall $\mu_1=15$, with fixed $\mu=1$ in the left panel and $\mu_2=12$ in the right panel.}
  \label{1}
\end{center}

\subsection{Exponential WTD - purely jump dynamics}
\label{sec:Exp2}
As already elaborated, in the considered case the time continuous
dynamics between the jumps does not strongly affect the qualitative
picture of non-Markovianity. It is therefore of interest to consider
the effect of jumps and modified waiting time distributions alone, setting $\mathcal{L}(t)=0$, see Eq.~\eqref{eq:RenGen}. In this case the density operator follows the evolution:
\begin{align}\label{noCont}
\rho(t)=\sum\limits_{n=0}^\infty p_n(t)\mathcal{E}^n \rho(0),
\end{align}
where $p_n(t)$ is the probability of having exactly $n$ jumps till
time $t$, i.e. no statements about the times of the particular jumps
are made as contrasted with $p_n(t; t_n,\ldots,t_1)$ in Eq.~\eqref{eq:RenGen}. As the influence of AD jump was also mainly in decreasing 
the non-Markovianity measure, with the same
argument we take $\mathcal{E} \rightarrow \mathcal{E}_x$ so that
we have idempotency of the jump transformation $\mathcal{E}^2=\id$. Accordingly, the sum in Eq.~\eqref{noCont} can be split in two terms, one with even $n$ and one with odd $n$, see \cite{Vacchini_2011} for an analogous discussion with the $z$-Pauli channel:
\begin{align}
\rho(t)=( p_{even}(t)+   p_{odd}(t)\mathcal{E})\rho(0).
\end{align}
The difference between the matrices $\rho_1(t)$ and $\rho_2(t)$ then
simply becomes
\begin{align}
\rho_1(t)-\rho_2(t)=\begin{pmatrix}
\Delta_{11}&  p_{even}(t) \Delta_{10} - p_{odd}(t) \Delta^*_{10}\\
p_{even}(t) \Delta^*_{10} - p_{odd}(t) \Delta_{10}\ & -\Delta_{11}
\end{pmatrix},
\end{align} 
where $\Delta_{ij}$ gives the difference of the associated components of the operators $\rho_1(t)$ and $\rho_2(t)$ in the $\sigma_z$ basis.
With the choice of the optimal states, $\ket{\phi_{1/2}}=\frac{1}{\sqrt{2}}(\ket{0} \pm i\ket{1})$, we obtain for the trace distance
\begin{align}
\label{pe-po}
\mathcal{D}(\rho_1(t),\rho_2(t))=| p_{even}(t)- p_{odd}(t)|=|q(t)|,
\end{align}
which is the absolute value of the difference between the probability of the even number of jumps and odd number of jumps. As distinct from investigations in \cite{Vacchini_2011}, here we take into account also case of modified processes, where first $k$ WTDs are different from the following one. The quantities $p_{even}(t)$ and $p_{odd}(t)$ take then in Laplace picture the form

\begin{align}
 \tilde{p}_{even}(u)&= \tilde{g}_1(u) +
                      \tilde{f}_1(u)\tilde{f}_2(u)\tilde{g}_3(u)+\ldots \nonumber  \\
                    &\hspace{1truecm}+\tilde{f}_1(u)\ldots\tilde{f}_k(u)
 \left(\frac {1+\tilde{f}(u)}2-(-)^k\frac {1-\tilde{f}(u)}2\right)
                     \frac{1}{1-\tilde{f}^2(u)} \tilde{g}(u), \label{eq1}\\
                     \tilde{p}_{odd}(u)&=\tilde{f}_1(u)\tilde{g}_2(u)+
                     \tilde{f}_1(u)\tilde{f}_2(u)\tilde{f}_3(u)\tilde{g}_4(u)+\ldots \nonumber \\
                    &\hspace{1truecm}+\tilde{f}_1(u)\ldots\tilde{f}_k(u)
\left(\frac {1+\tilde{f}(u)}2+(-)^k\frac {1-\tilde{f}(u)}2\right)
                    \frac{1}{1-\tilde{f}^2(u)} \tilde{g}(u). \label{eq2}
\end{align}
In the case of the exponential WTDs we can accordingly go beyond the
Markovian case of an exponential distribution corresponding to $q(t)=e^{-2 \mu t}$. For the simplest case of 2 WTDs one obtains
\begin{align}
q(t)&=\frac{2 (\mu -\mu_1) e^{ -t \mu_1}+\mu_1 e^{-2 \mu  t}}{2 \mu -\mu_1}.
\end{align}  
The expression of $q(t)$ for a larger number of WTDs retains the same form, i.e. a weighted sum of $k$ exponentials $e^{-2 t \mu }$, $e^{-t \mu _1}$,\ldots, $e^{-t \mu _k}$.
Non monotonicity of the absolute value of the function $q(t)$ can
arise in two ways: non monotonicity of $q(t)$ itself or its
sign change. Note that these are not independent, as $q(t)$
convergences to zero for $t \rightarrow \infty$. Accordingly, with
every sign change at least one local maximum or minimum has to
follow. On the other hand, a local maximum (minimum) can occur without
sign change, but then need to be followed by a minimum (maximum).

For the case of two waiting time distributions one can analytically
verify that the maximal number of revivals is one, and that revivals take place at 
\begin{align}
t=-\frac{1}{2\mu-\mu_1}\ln\frac{2(\mu_1-\mu)}{\mu_1},
\end{align}  
where the condition $\mu_1>\mu$ has to be satisfied. This corresponds to
the requirement obtained for the dynamics considered in Sect.~\ref{sec:Exp1}, Eq.~\eqref{relation},
which, however, could feature an intermediate time continuous evolution and a
jump transformation containing AD.
Note that the time $t$ is smaller than the mean jump time of the first jump ${1}/{\mu_1}$ for $\mu_1>2\mu$, otherwise it is larger. For larger $k$ in general no closed-form formula for the number or the times of revivals can be given, as the exponential function is transcendental. Nonetheless, thanks to the Descartes’ rule of signs the maximal number of revivals ($q(t)=0$) equals the number of the sign changes of the prefactors of the exponential functions, where the rates are put in ascending (or descending) order \cite{10.2307/40378610}. The sign change can happen maximally $k-1$ times for $k$-terms, which explains the observation we have made earlier in Sect.~\ref{sec:Exp1}. Note, that the same argument could be used for the derivative of $q(t)$, connected with the occurrence of local maxima/minima. However, the maximal number of revivals $k-1$ can only happen when all of the revivals are at zero distance, as the non-monotonicity of $q(t)$ without sign change involves one minimum and one maximum per revival. Note that consequently for a process with 2 WTDs the revival can only occur because of the sign change of $q(t)$, i.e. at zero trace distance.  

\subsection{Erlang WTD - general results}
\label{sec:Erl1}

Considering WTDs that can lead to non-Markovianity for unmodified
processes, the maximal possible number of revivals can get larger. This
can be observed by taking into account an Erlang distribution, whose WTD is given by Eq.~\eqref{erlang}, governing the
randomness of the jump times. For Erlang distributions with fixed mean
value, the higher the shape parameter $r$ or the larger the rate $\mu$, the
narrower the distribution. Accordingly, with growing $r$ or $\mu$ the revivals of the trace distance
can be seen more and more like independent phenomena. In this case the
jumps do not "destructively interfere" with each other and the time
intervals of the jumps are almost disjoint. This explains the increase
of the non-Markovianity measure with higher shape parameter $r$ or larger
rate $\mu$, as one can see in the simulations in Fig.~\ref{Erl1}, left. {This slightly influences also the time of the revivals, and the higher the shape parameter, the closer this time is to the mean value of the first WTD.}    

  \begin{center}
       \minipage{0.3\textwidth}
    \includegraphics[width=\textwidth]{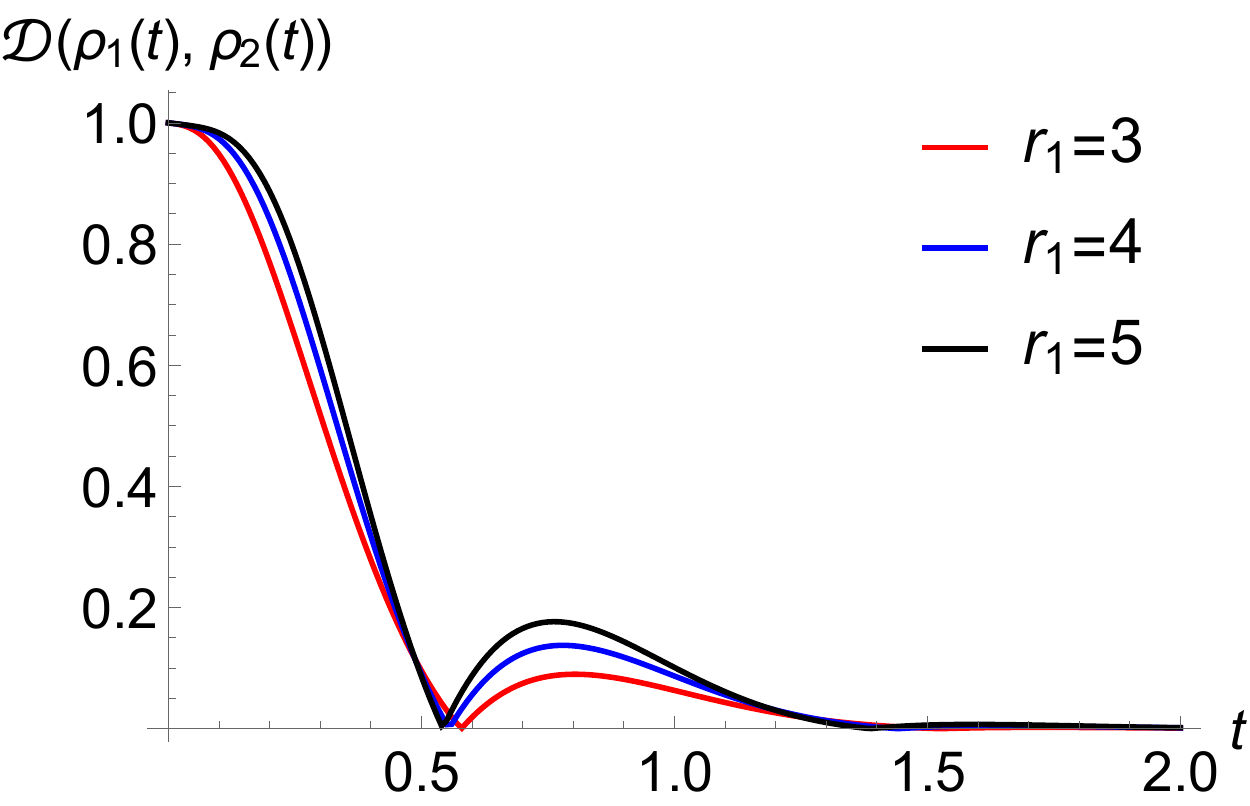}
    \endminipage
    \hspace{0.5cm}
         \minipage{0.3\textwidth}
      \includegraphics[width=\textwidth]{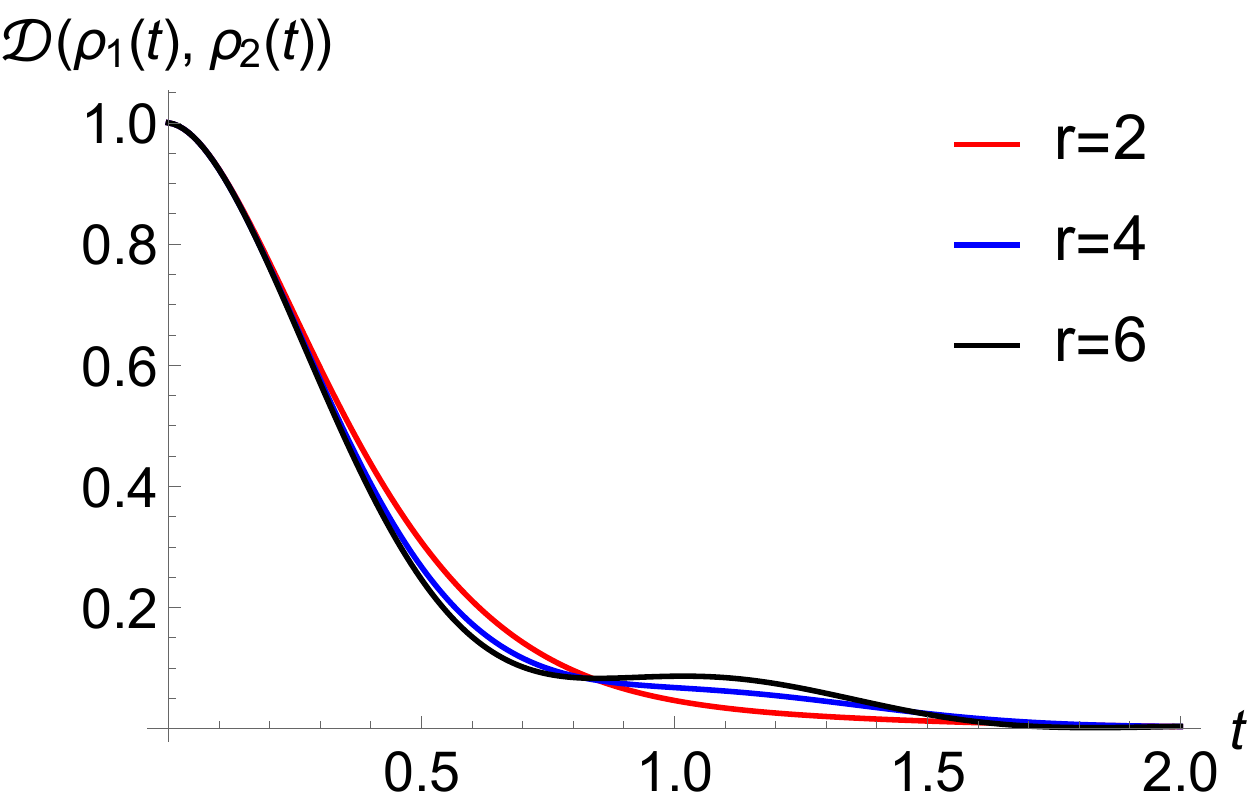}  
      \endminipage
      \hspace{0.5cm}
           \minipage{0.3\textwidth}
      \includegraphics[width=\textwidth]{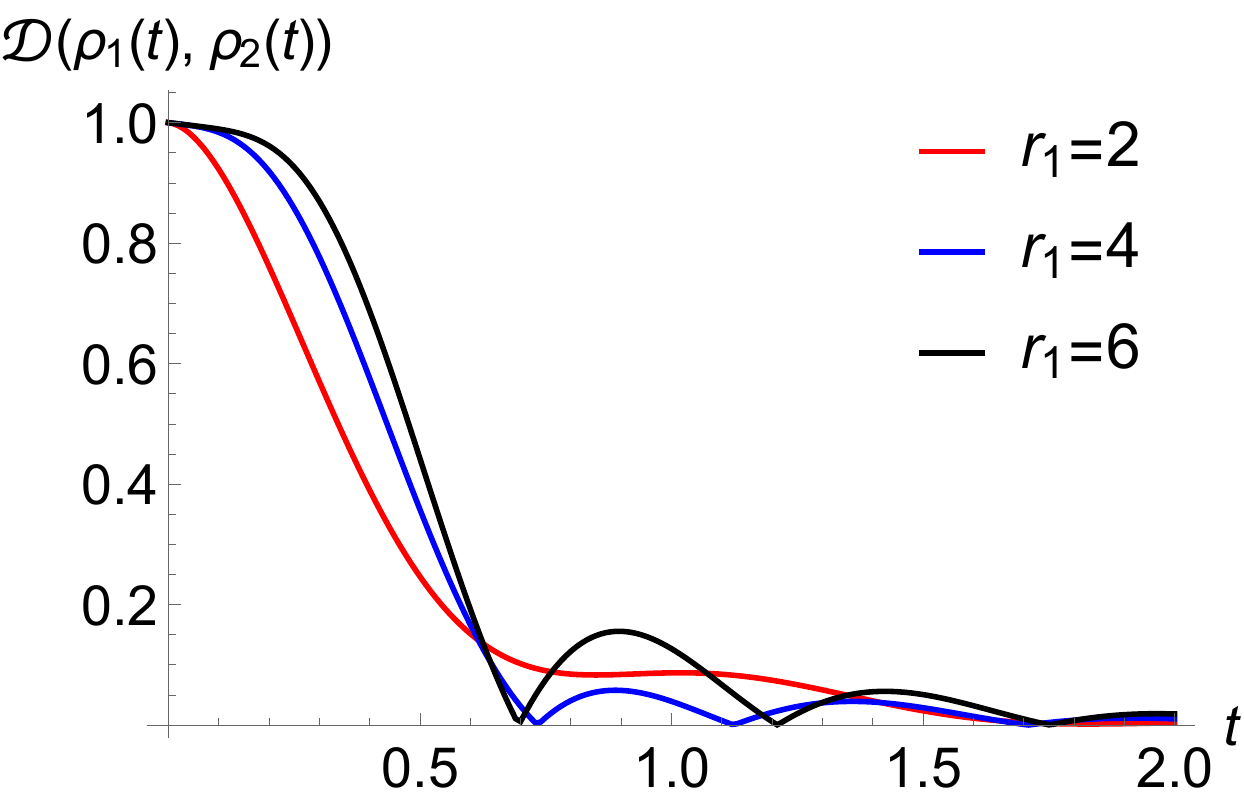}
  \endminipage
  \captionof{figure}{Behavior of the trace distance for the case of a quantum renewal process with a $\mathcal{E}_{x-\mathrm{AD}}$ jump channel and WTDs given by Erlang distributions. In the left and right panel we see that revivals increase with the shape parameter $r_1$ ($\mu=4$, $r=2$, $r_1/\mu_1=1/2$ left and $\mu=12$, $r=6$, $r_1/\mu_1=2/3$ right). In the middle panel we see dependence on the parameter $r$ with fixed
  $\mu_1=3$, $r_1=2$ and $r/\mu=1/2$.}
  \label{Erl1}
\end{center}

Also in the case of the Erlang WTDs the trace distance revivals do not
necessarily occur when the trace distance takes the value zero. This
behaviour was observed for modified renewal processes. The small $r$ of the first WTD and the large $r$ of the subsequent WTD
boost the phenomenon, see Fig.~\ref{Erl1}, middle and right. 
Note, that contrary to the case of the
exponential WTDs, here the revival can occur at non-zero trace
distance also for the simplest modified process, i.e. with 2 distinct
WTDs.

\subsection{Erlang WTD - purely jump dynamics}
\label{sec:Erl2}

For the limiting case of no time continuous evolution in between the jumps $\mathcal{E}_{x}$, relying on Eq.~\eqref{pe-po} for the trace distance between the optimal pair of states, one can analytically show that an infinite number of revivals is possible. The difference of the probability of the even and odd number of jumps for an unmodified process is given in Laplace domain by
\begin{align}
\tilde{q}(u)=\frac{(\mu +u)^r-\mu ^r}{u \left(\mu ^r+(\mu +u)^r\right)}.
\end{align}
In particular, for $r=2$, so for WTD given by a convolution of two exponential functions with the same rate, we obtain
\begin{align}
q(t)=e^{-\mu t }(\sin(\mu t) +\cos(\mu t) ),
\end{align}
which obviously leads to an infinite number of revivals, always occurring at the zero trace distance.
For the modified process, with two different WTD and when both shape parameters equal two, $r=r_1=2$, one obtains 
\begin{align}
q(t)=\frac{1}{\left(2 \mu ^2-2 \mu _1 \mu +\mu _1^2\right){}^2} 
\left(2 \left(\mu _1-\mu \right) e^{-\mu_1  t} \left(\mu _1^3-3 \mu _1^2 \mu
+2 \mu _1 \mu ^2 -2 \mu ^3
+t\mu _1 \left(\mu _1^3-3 \mu _1^2 \mu
+4 \mu _1 \mu ^2 -2 \mu ^3\right)
\right) \right.
\nonumber \\ \left.
-\mu _1^2 e^{-\mu  t} \left( ((2\mu-\mu_1)^2-2\mu^2) \cos(\mu t) - (2\mu^2-\mu_1^2)  \sin(\mu t) \right)\right).\label{eq:qErl}
\end{align}
Accordingly, we have a term characterised by an oscillation, which is damped with a damping rate $\mu$, and a polynomial of the first order in $t$, damped with a damping rate $\mu_1$. From Fig.~\ref{Erl2} we see that for $\mu=1$ and $r=r_1=2$, if the rate of the first waiting time distribution $\mu_1$ is between zero and a value close to one, no revivals take place. This can be understood from Eq.~\eqref{eq:qErl}, since    
if the rate $\mu$ is larger than the rate of the first WTD, the oscillatory part is strongly suppressed. However, for this regime the polynomial part stays always positive, and no revivals occur. Otherwise, we have an infinite number of revivals.

  \begin{center}
       \minipage{0.4\textwidth}
    \includegraphics[width=\textwidth]{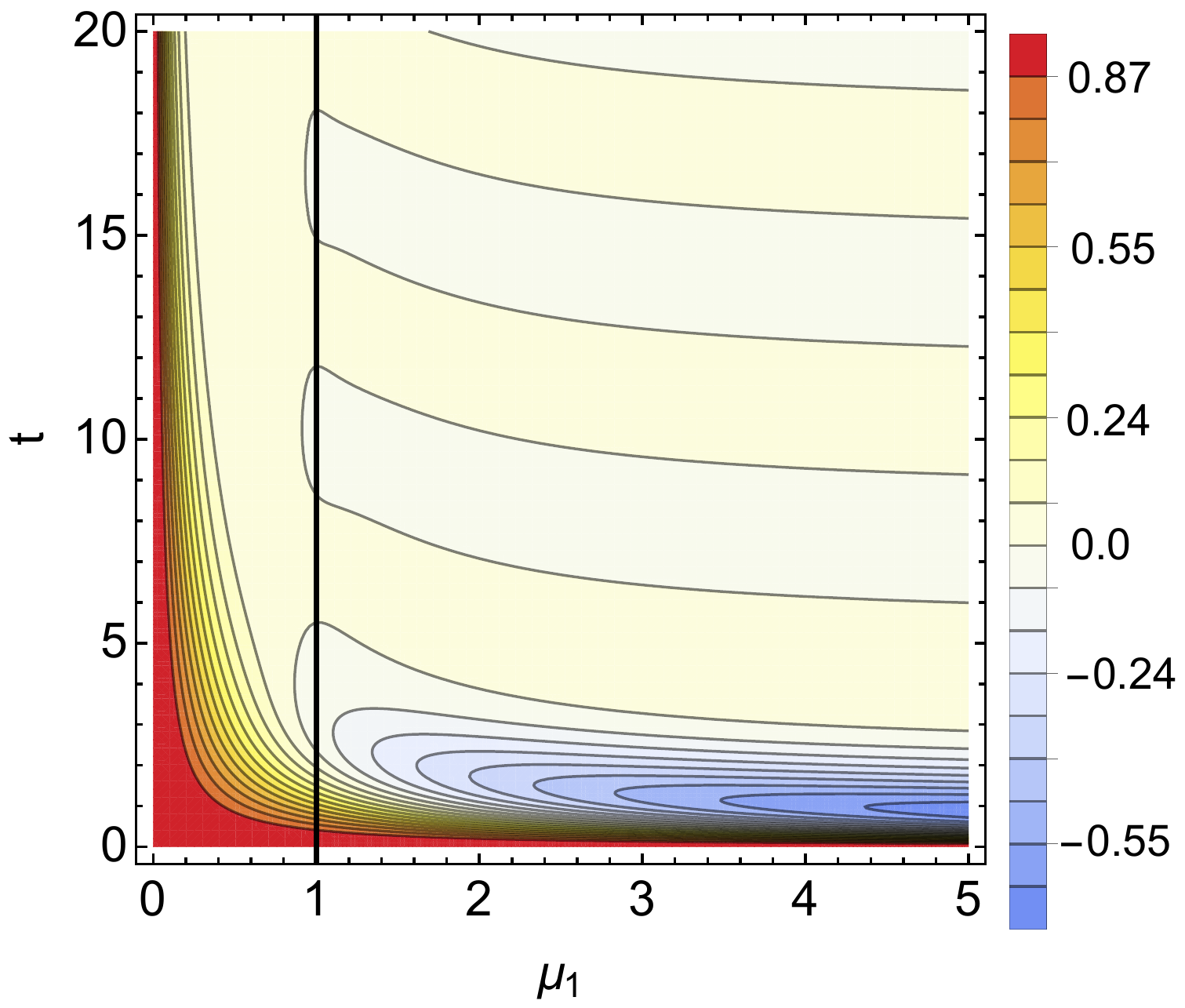}
  \endminipage
  \captionof{figure}{Plot of the function $q(t)$ as in Eq.~\eqref{eq:qErl}. The non-monotonicity of this function determines non-Markovianity in the model, see Eq.~\eqref{pe-po}.  The function $q(t)$ corresponds to the difference between the probability of having an even or an odd number of jumps as a function of time and WTDs' rates. We consider two Erlang WTDs with shape parameters $r=r_1=2$ and $\mu=1$, so that the vertical black line corresponds an unmodified renewal process.  Note the periodic change of values along the vertical axis determining an infinite number of revivals.}
  \label{Erl2}
\end{center}

\section{Conclusions and outlooks}
\label{sec:concl}

In this work, 
we have analysed a simple and versatile class of quantum dynamics, the quantum renewal processes, 
focusing on the different kinds of non-Markovian behavior that can be obtained by controlling their 
defining properties.

Quantum renewal processes naturally allow for a representation of the dynamics in terms of an average over stochastic trajectories
and we have here investigated the influence that the time-continuous part of the trajectories, the type of the jumps and the waiting time distributions have on the quantitative and qualitative features of the trace distance evolution. 
In particular, we focused not only on the measure of non-Markovianity, but also on relevant modifications of the trace distance
evolution, as the number, times of occurrence and extension of the revivals. 
Among others, the revivals of the trace distance
can be significantly altered or even
enhanced when dealing with
modified renewal processes, where there is a difference between a certain number of initial waiting time distributions and the subsequent ones, 
or if one considers Erlang waiting time distributions, which are classically non-Markovian and can lead to higher number of revivals than the exponential ones.

Our analysis shows that 
the trajectory picture of quantum renewal processes yields a deeper insight into how to manipulate the trace distance
evolution,
for a varied class of
dynamics built on the analogy with classical stochastic processes.
Indeed, it will be of interest to explore to which extent the trajectory viewpoint can be a convenient starting
point to engineer non-Markovianity also in more complex and general quantum dynamics, pointing to different
features of the evolution that can be addressed to enhance or suppress the 
presence of
memory effects.






\begin{acknowledgments}
{This research was funded by the UniMi Transition Grant H2020. N.M. was funded by the Alexander von Humboldt Foundation in form of a Feodor-Lynen Fellowship.}
\end{acknowledgments}

\bibliographystyle{unsrt}
\bibliography{QRP}

\end{document}